\documentclass[%
aps,
pra,
reprint,
nofootinbib,
amsmath,
amssymb,
longbibliography
]{revtex4-1}
\usepackage{
physics,
graphicx,
booktabs,
mathtools, 
placeins, 
hyperref 
}
\usepackage{adjustbox}
\usepackage[dvipsnames]{xcolor}
\hypersetup{
    colorlinks,
    linkcolor={blue!50!black},
    citecolor={blue!50!black},
    urlcolor={blue!80!black}
}

\usepackage[caption=false]{subfig} 
\usepackage[capitalize]{cleveref}
\usepackage{tikz}
\usetikzlibrary{quantikz}

\renewcommand{\var}[1]{{\mathrm{Var}\left[#1\right]}}

\begin{document}


\title{Resource Estimation for Quantum Variational Simulations of the Hubbard Model}

\author{Zhenyu Cai}
\email{cai.zhenyu.physics@gmail.com}
\affiliation{Department of Materials, University of Oxford, Oxford, OX1 3PH, United Kingdom}
\affiliation{Quantum Motion Technologies Ltd, Nexus, Discovery Way, Leeds, LS2 3AA, United Kingdom}

\date{\today}

\begin{abstract}
    As the advances in quantum hardware bring us into the noisy intermediate-scale quantum (NISQ) era, one possible task we can perform without quantum error correction using NISQ machines is the variational quantum eigensolver (VQE) due to its shallow depth. A specific problem that we can tackle is the strongly interacting Fermi-Hubbard model, which is classically intractable and has practical implications in areas like superconductivity. In this Article, we outline the details about the gate sequence, the measurement scheme and the relevant error mitigation techniques for the implementation of the Hubbard VQE on a NISQ platform. We perform resource estimation for both silicon spin qubits and superconducting qubits for a 50-qubit simulation, which cannot be solved exactly via classical means, and find similar results. The number of two-qubit gates required is on the order of $20000$. Hence, to suppress the mean circuit error count to a level such that we can obtain meaningful results with the aid of error mitigation, we need to achieve a two-qubit gate error rate of $\sim 10^{-4}$. When searching for the ground state, we need a few days for one gradient-descent iteration, which is impractical. This can be reduced to around $10$ minutes if we distribute our task among hundreds of quantum processing units. Hence, implementing a 50-qubit Hubbard model VQE on a NISQ machine can be on the brink of being feasible in near term, but further optimisation of our simulation scheme, improvements in the gate fidelity, improvements in the optimisation scheme and advances in the error mitigation techniques are needed to overcome the remaining obstacles. The scalability of the hardware platform is also essential to overcome the runtime issue via parallelisation, which can be done on one single silicon multi-core processor or across multiple superconducting processors. 
\end{abstract}

\maketitle

\section{Introduction}
The difficulty of simulating large quantum systems was an inspiration for the idea of quantum computation~\cite{feynmanSimulatingPhysicsComputers1982}, thus naturally, we look to quantum system simulation as one of the first applications of quantum computers. Fault-tolerant simulation of a non-trivial system requires an integrated quantum device with at least hundreds of thousands of qubits given a gate infidelity of $10^{-3}$ or tens of thousands of qubits given a gate infidelity of $10^{-4}$~\cite{kivlichanImprovedFaultTolerantQuantum2019}. This is still out of reach with the current technology. With the advance of qubit counts and gate error rates bringing us into the noisy intermediate-scale quantum (NISQ) era, one must wonder if there is a classically intractable and physically meaningful task such NISQ hardware can tackle without quantum error correction. One of the most promising candidates is the variational quantum eigensolver (VQE) due to its shallow depth. In VQE, we aim to use a parametrised quantum circuit to prepare the eigenstate (usually the ground state) of a given Hamiltonian. The task is carried out through measuring observables using the quantum circuit and optimising the parameters using a classical computer. As we will see later, one of the lowest hanging fruits in this area is the preparation of the ground state of the strongly interacting Fermi-Hubbard model, which is what we will be focusing on in this Article. 

A huge number of circuit runs are needed to run VQE~\cite{weckerProgressPracticalQuantum2015}. Luckily, the task is highly parallelisable and thus can be distributed to many quantum processors. Both silicon spin qubits and superconducting qubits are good candidates for such a task due to their compatibility with the commercial fabrication technology~\cite{vandersypenInterfacingSpinQubits2017, kjaergaardSuperconductingQubitsCurrent2020}, which should reduce the cost and enhance the reliability of reproducing multiple quantum processors for the same task. Furthermore, we presently explain that the ansatz circuit of the Hubbard model that we are interested in can be naturally broken into the native gates of both silicon spin qubits and superconducting qubits. In this Article, we will first use silicon spin qubits as our example platform to carry out our gate count and runtime analysis and then we will apply the same analysis to the superconducting qubits. 

The Article is structured as follows, we begin by introducing the concept of VQE and why we choose to simulate the Hubbard model in \cref{sect:vqe}. Then we will outline the ansatz circuit we use and its gate counts for the silicon platform in \cref{sect:hubbard_ansatz}. To deal with the noise in the circuit, we will apply error mitigation to our example circuit in \cref{sect:error_mitigation}. Following that, in \cref{sect:runtime_est} we will introduce more details about our implementation and the estimated algorithm runtime on the silicon platform. In \cref{sect:resource_est_super}, we will apply the same analysis to the superconducting platform. This is followed by our conclusion in \cref{sect:conclusion}.

\quad\\

\section{Variational Quantum Eigensolver}\label{sect:vqe}
\subsection{Background}
For a given Hamiltonian $H$, we want to find a circuit that can prepare its ground state $\ket{\psi_0}$ with the associated ground state energy $E_0$. We try to achieve this with a circuit with tunable gates that have $M$ parameters $\vec{\theta} = \{\theta_1, \theta_2, \cdots, \theta_M\}$ that we can control, which will produce an output state $\ket{\psi(\vec{\theta})}$. In VQE, Our goal is to obtain the set of optimal parameters $\vec{\theta}_{op}$ such that the state prepared by the circuit can well approximate the ground state we want $\ket{\psi(\vec{\theta}_{op})} \approx \ket{\psi_0}$. Then we can measure various properties of the ground state using $\ket{\psi(\vec{\theta}_{op})}$.
Refs~\cite{mcardleQuantumComputationalChemistry2020} and \cite{caoQuantumChemistryAge2019} provide a comprehensive overview on variational algorithms and more generally the field of quantum computational chemistry. 
\subsection{Choosing the Simulation Problem and its Corresponding Ansatz}\label{sect:choosing problem}
The parametrised circuit in VQE is called an ansatz circuit, or simply ansatz for short. It determines the quantum subspace that our output state can reach, hence it is the key to the success of our algorithm. An example of a simple ansatz is called the hardware-efficient ansatz (HEA)~\cite{kandalaHardwareefficientVariationalQuantum2017}, which just uses gates available to the physical qubit systems to create many entangling blocks along the circuit. However, it is difficult to obtain an analytical or even a heuristic estimate of the number of gates required in an HEA for the success of our algorithm. A better approach will be using an ansatz inspired by the problem itself. For example, the unitary-coupled-cluster ansatz (UCCA) for quantum chemistry~\cite{romeroStrategiesQuantumComputing2018}, the low depth circuit ansatz (LDCA)~\cite{dallaire-demersLowdepthCircuitAnsatz2019} and the Hamiltonian ansatz (HA)~\cite{weckerProgressPracticalQuantum2015} for more general simulations of closed quantum systems. 

The number of gates needed by HA scales as $\mathcal{O}(N_{blk}R)$ where $R$ is the number of subterms in the Hamiltonian of our simulation and $N_{blk}$ is the number of repeating blocks within the ansatz. For a general electronic structure Hamiltonian (which applies to most chemistry Hamiltonians), we have $R \sim \mathcal{O}(N^4)$ due to the terms accounting for electron-electron interactions~\cite{weckerProgressPracticalQuantum2015}. For periodic systems, the added structure allows us to transform our basis orbitals to achieve $R \sim \mathcal{O}(N^2)$~\cite{babbushLowDepthQuantumSimulation2018}.  One of the systems that has the most favourable scaling while maintaining great physical interest is the 2D Hubbard model, in which by restricting to only on-site interactions and nearest-neighbour hopping, we can achieve $R \sim \mathcal{O}(N)$. Using HA to simulate the Hubbard model, the gates we apply are dependent on the qubit encoding we use. Using the Jordan-Wigner encoding, some of the gates will be non-local, which can be overcome by using $\mathcal{O}(N^{\frac{1}{2}})$ additional gates~\cite{kivlichanQuantumSimulationElectronic2018}. Other encodings like Verstraete-Cirac encoding~\cite{verstraeteMappingLocalHamiltonians2005} or superfast encoding~\cite{bravyiFermionicQuantumComputation2002} will ensure locality, but require at least doubling the number of qubits. Since we are focusing on near-term devices in which qubit resources can be limited, we will only consider the Jordan-Wigner encoding in this Article. In such case, the number of gates needed to simulate the 2D Hubbard Model using HA scales as $N_{gates} \sim \mathcal{O}(N_{blk}N^{\frac{3}{2}})$. This is a factor of $N^{\frac{1}{2}}$ better than LDCA assuming the same number of blocks $N_{blk}$ in both ansatze. To achieve results of sufficient precision, we likely need to go up to double excitation for UCCA, which means a gate scaling of $N_{gates} \sim \mathcal{O}(N^5)$ (assuming Jordan-Wigner encoding, see Ref~\cite{caoQuantumChemistryAge2019}). To achieve similar precision, $N_{blk}$ for HA is likely to scale better than $\mathcal{O}(N^{\frac{7}{2}})$ based on the previous Hubbard model numerical simulation results~\cite{weckerProgressPracticalQuantum2015,reinerFindingGroundState2019,cadeStrategiesSolvingFermiHubbard2019}, thus 
HA should have a scaling advantage over UCCA in simulating the Hubbard model.
 
Hence, in the rest of this Article, we will be focusing on the ground state preparation of the Hubbard model using Hamiltonian Ansatz (HA) under the Jordan-Wigner encoding. 

Note that for the NISQ regime that we are interested in, the constants we ignored in the above scaling analysis can make a very significant difference. Thus there might exist other problems with a different ansatz implementation that are more practical than the one we are going to consider when we dig into the exact implementation details.  

\section{Ansatz Circuit}\label{sect:hubbard_ansatz}
\subsection{2D Fermi-Hubbard Model}\label{sect:hubbard_model}
The Hamiltonian of the 2D Fermi-Hubbard model is:
\begin{align*} 
H = -t \sum_{\sigma, \expval{v, w}} \left(a^\dagger_{v, \sigma} a_{w, \sigma} + a^\dagger_{w, \sigma} a_{v, \sigma}\right) + U \sum_v n_{v, \uparrow}  n_{v, \downarrow}
\end{align*}
where $a^\dagger_{v, \sigma}/a_{v, \sigma}$ are the creation/annihilation operators of site $v$ with spin $\sigma$. The first term represents the nearest neighbour hopping interactions, with $t$ being the tunnelling energy and $\expval{v, w}$ representing summing up sites $v$ and $w$ that are adjacent to each other in a 2D geometry. The second term is the on-site repulsion energy. 

Extracting the ground state properties of the Hubbard model in the parameter regime of e.g. $\frac{U}{t} \approx 4 \rightarrow 8$ at close to half-filling is believed to be relevant to the understanding of high-$T_c$ cuprate superconductors~\cite{simonscollaborationonthemany-electronproblemSolutionsTwoDimensionalHubbard2015,weckerSolvingStronglyCorrelated2015}. However, solutions using classical methods have high degree of uncertainty in this parameter regime ~\cite{simonscollaborationonthemany-electronproblemSolutionsTwoDimensionalHubbard2015}, leading to our attempt here using quantum algorithms. In this Article, for simplicity we will be considering the half-filling case, i.e. for $V$ sites we have $V$ electrons, but most of our arguments can be generalised to any number of electrons.

Using the Jordan-Wigner encoding, each qubit will encode one orbital. There are two spin orbitals to each site, hence the number of qubits ($N$) needed is twice the number of sites ($V$). For a classically intractable $50$-qubit problem, we will be looking at $V = 25$, e.g. a $5\times5$ Hubbard model. It is important to note that even though a Hubbard Hamiltonian of this size cannot be solved exactly classically, there are various approximation methods available~\cite{simonscollaborationonthemany-electronproblemSolutionsTwoDimensionalHubbard2015}.

In the Jordan-Wigner encoding, the Fermionic creation and annihilation operators become:
\begin{align*}
a_i^\dagger &\mapsto Z_{:i} A_i^\dagger\\
a_i &\mapsto Z_{:i} A_i
\end{align*}
where $i$ denotes the index of the canonical ordering of the orbitals (including both spins and sites). $A = \ket{0}\bra{1} = \frac{X + iY}{2}$ is the qubit lowering operator and $A^\dagger$ is the qubit raising operator. $Z_{:i} = \prod_{j=1}^{i-1} Z_j$ is the $Z$ operator string that maintains the Fermionic commutation relationship.

The on-site repulsion term and the hopping term of the Hubbard Hamiltonian thus becomes:
\begin{equation}
\begin{split}\label{eqn:jw_terms}
n_in_j = a^\dagger_ia_i a^\dagger_ja_j &\mapsto \frac{1}{4} (I - Z_i)(I - Z_j)\\
a_i^\dagger a_{j} + a_{j}^\dagger a_{i}&\mapsto \frac{1}{2}\left(X_iX_{j} + Y_iY_{j}\right) Z_{i:j}
\end{split}
\end{equation}
where we have without loss of generality assumed $j>i$ and defined $Z_{i:j} = \prod_{k=i+1}^{j-1} Z_k$.

\subsection{Hamiltonian Ansatz}
The Hamiltonian ansatz was proposed by Wecker~\textit{et al.}~\cite{weckerProgressPracticalQuantum2015}. Its circuit is essentially a Trotterised variational annealing path, which is inspired by both adiabatic state preparation and the quantum approximate optimisation algorithm~\cite{farhiQuantumApproximateOptimization2014, farhiQuantumSupremacyQuantum2016}. 

For a Hamiltonian of the form:
\begin{align*}
H = \sum_i \lambda_i h_i
\end{align*}
where $h_i$ are different interaction terms, a block of the Hamiltonian ansatz is just:
\begin{align*}
\prod_i e^{-i\theta_i h_i}
\end{align*}
i.e. we implement a parametrised unitary gate corresponding to every interaction term. We will implement $N_{blk}$ of such blocks in sequence with different sets of parameters. The Hamiltonian ansatz, though inspired by Trotterised annealing, is not equivalent to Trotterisation. Thus a Hamiltonian ansatz using higher-order Trotterisation does not necessarily have lower algorithmic errors than a Hamiltonian ansatz using lower-order Trotterisation. In NISQ devices, we might want to avoid using higher-order Trotterisation in the Hamiltonian ansatz due to higher gate counts and deeper circuits. Hence, here we have used the first-order Trotter formula for the Hamiltonian ansatz. 

From \cref{eqn:jw_terms}, we see that the gates corresponding to the repulsion terms are all local, while the gates that correspond to hopping will only be local if the two orbitals involved are close to each other in the canonical ordering due to the trailing $Z$ string.

\begin{figure}[htbp]
    \centering
    \includegraphics[width = 0.4\textwidth]{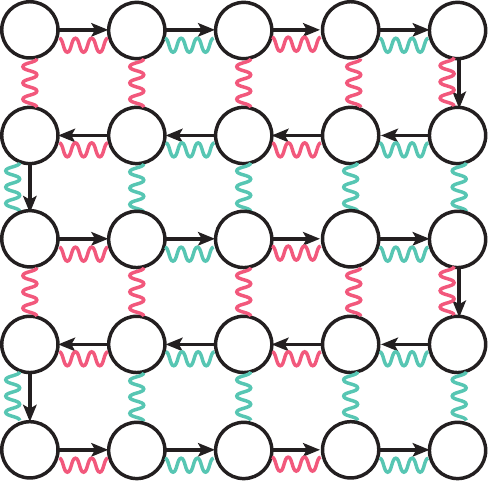}
    \caption{The black arrows denote the canonical order of the orbitals of different sites for a $5\times 5$ Hubbard model. Here the ordering is in a snake-fold pattern running along the horizontal direction. The red/blue curly lines denote the even/odd hopping interaction.}
    \label{fig:site_order}
\end{figure}

Now if we choose a canonical ordering such that the two spin orbitals of the same site are always adjacent to each other, while the orbitals of different sites are ordered in a folding pattern running along the horizontal direction as shown in \cref{fig:site_order}, then all the horizontal hopping terms will be local while some of the vertical hopping terms will not be. The non-local vertical hopping can be made local by using additional Fermionic swap (fSWAP) gates to swap the canonical order of the orbitals~\cite{verstraeteQuantumCircuitsStrongly2009}:
\begin{align*}
f_{\text{swap}}^{i, i+1} & \mapsto \frac{1}{2}\left[ \left(X_iX_{i+1} + Y_iY_{i+1}\right) + Z_i + Z_{i+1}\right].
\end{align*}

Kivlichan~\textit{et al.}~\cite{kivlichanQuantumSimulationElectronic2018} outline a way to implement a block of the Hamiltonian ansatz for the open boundary 2D Hubbard model using only local gates with $\mathcal{O}(N^{\frac{1}{2}})$ depth using a Fermionic swap network. The exact structure of this ansatz is recapped in \cref{sect:sim_scheme}. In \cref{sect:ansatz_circuit}, we have decomposed the gates corresponding to the repulsion terms, the hopping terms and the Fermionic swap into the native gates of silicon spin qubits, and found that we only need single-qubit $Z$ rotation and partial swaps. Then we applied the gate decompositions to the Hamiltonian ansatz, optimised the circuit using the qubit exchange symmetry of the gates, and obtained the number of single-qubit gates and two-qubit gates needed for one block of Hamiltonian ansatz for $V$ sites to be:
\begin{align*}
N_{1q, ha} & = 4V^{\frac{3}{2}} + 7V - 4\sqrt{V}\\
N_{2q, ha} & = 8V^{\frac{3}{2}} + V - 4\sqrt{V}.
\end{align*}
For $V = 25$, we have
\begin{align*}
N_{1q, ha} &\approx 650\\
N_{2q, ha} &\approx 1000.
\end{align*}

\subsection{Full Ansatz Circuit}\label{sect:gate_count}
To produce a good approximation to the Hubbard ground state, we need a good starting state for the Hamiltonian ansatz. In the context of adiabatic evolution, a good starting state should be in the same phase as the output state as discussed by Wecker \textit{et al.}~\cite{weckerSolvingStronglyCorrelated2015}. Since the Hamiltonian ansatz evolves out of adiabatic evolution, we should expect a similar result to hold. The starting state we use should be a single Slater determinant, which can be solved classically and can be efficiently prepared using a quantum circuit~\cite{jiangQuantumAlgorithmsSimulate2018, kivlichanQuantumSimulationElectronic2018}, e.g. the ground state of the non-interacting Hubbard Hamiltonian (i.e. $U=0$). In \cref{sect:state_prep} we have recapped the details about the Slater determinant preparation circuit using Givens rotation. We decomposed the Givens rotation into the native gates of silicon spin qubits and again found that we only need $Z$ rotation and partial swaps. Then we applied the gate decompositions to the Slater determinant preparation circuit, optimised the circuit and obtained the number of single-qubit gates and two-qubit gates needed for $V$ sites to be:
\begin{align*}
N_{1q, prep} &= 2V^2\\
N_{2q, prep} &= 2V^2.
\end{align*}
When the two spin subspaces of the starting state are decoupled (e.g. the non-interacting Hubbard ground state), we can prepare the Slater determinant separately in the two spin subspaces, which is also discussed in \cref{sect:state_prep}. In this case, we need to start in a orbital ordering in which the spin-up and spin-down are separated, and end in a orbital ordering in which the spin-up and spin-down are interleaved for inputting into the Hamiltonian ansatz. We found that the saving in gate counts is limited due to the need to rearrange the orbital ordering while the runtime needed is longer. Hence, here we will stick with the simple Slater determinant preparation scheme in which we do not consider the two spin subspaces separately. 

The total number of gates needed for the whole ansatz circuit on the silicon-spin-qubit platform is:
\begin{align*}
N_{1q} &= \underbrace{2V^2}_{N_{1q, prep}} + \underbrace{\left(4V^{\frac{3}{2}} + 7V - 4\sqrt{V}\right)}_{N_{1q, ha}}N_{blk}\\
N_{2q} &= \underbrace{2V^2}_{N_{2q, prep}} + \underbrace{\left(8V^{\frac{3}{2}} + V - 4\sqrt{V}\right)}_{N_{2q, ha}} N_{blk}.
\end{align*}
Hence, if $N_{blk}$ scales better than $\mathcal{O}(\sqrt{V})$, then the Slater determinant preparation will dominate the gate count at large $V$, otherwise the Hamiltonian ansatz will dominate the gate count at large $V$. In the previous simulation of the Hubbard model in a ladder grid structure with periodic boundaries~\cite{weckerProgressPracticalQuantum2015}, $N_{blk}$ scales super-linear to $V$, while Ref~\cite{reinerFindingGroundState2019, cadeStrategiesSolvingFermiHubbard2019} also shows similar results for the open-boundary Hubbard model. Hence, we will expect the gate cost due to the Slater determinant preparation part of the circuit to be negligible at large $V$.

If we take the optimistic assumption of $N_{blk} \sim V$, then the number of gates needed in the ansatz circuit for $V = 25$ on the silicon-spin-qubit platform will be:
\begin{equation}\label{eqn:gate_count_25}
\begin{split}
N_{1q} &\approx 17000\\
N_{2q} &\approx 26000.
\end{split}
\end{equation}
Across all major qubit platforms, the error rate of the single-qubit gates is generally much lower than that of the two-qubit gates. If we assume an optimistic two-qubit gate error rate of around $10^{-4}$ and a negligible single-qubit gate error rate, the expected number of errors in each circuit run (which we will call the \emph{mean circuit error count} from here on) is: 
\begin{align}\label{eqn:circ_err_rate}
\mu = 26000 \times 10^{-4} \sim 2.5,
\end{align}
which is still of the order of unity. Hence, to obtain a meaningful result out of our noisy circuit, we must apply error mitigation.

\section{Error Mitigation}\label{sect:error_mitigation}
VQE is inherently robust against local systematic errors since they can be offset by shifts in the variational parameters~\cite{mccleanTheoryVariationalHybrid2016}, which was observed in experiments~\cite{omalleyScalableQuantumSimulation2016}. For the other error components, we can further mitigate them via symmetry verification.

\subsection{Symmetry Verification}\label{sect:sym_ver}
It is often the case that there are some symmetries that the resultant state must follow. Suppose we want to obtain the expectation value of an observable $O$ using the output state, and we know that our output state must follow the symmetry $S$. We can measure both $O$ and $S$ in every circuit run and discard the runs that fail the symmetry test to obtain the error-mitigated expectation value~\cite{bonet-monroigLowcostErrorMitigation2018, mcardleErrorMitigatedDigitalQuantum2019}, which is called \emph{direct verification}. Suppose the fraction of circuit runs that passed the symmetry test is $P_S$, then we need a factor of 
\begin{align}\label{eqn:sym_cost}
C_{S} = \frac{1}{P_S}
\end{align}
more circuit runs to implement direct verification. $C_{S}$ here is the sampling cost factor of direct verification.

In the case that $O$ and $S$ cannot be measured in the same run and $S$ is a Pauli symmetry with eigenvalues $s = \pm 1$, the symmetry-verified expectation value $\overline{O}_{S}$ can be obtained via post-processing instead~\cite{bonet-monroigLowcostErrorMitigation2018}:
\begin{align}\label{eqn:sym_verification}
\overline{O}_{S} = \frac{\overline{O} + s \overline{OS}}{1 + s \overline{S}}.
\end{align}
Here we have assumed $O$ commute with $S$, which is often the case when $O$ is a term in the Hamiltonian since many symmetries of the ground state follow from the symmetries of the Hamiltonian. As shown in Ref.~\cite{hugginsEfficientNoiseResilient2019}, the cost of post-processing verification will be the square of that of direct verification.

\subsection{Symmetry Verification in Hubbard Model Simulation}\label{sec:sym_ver_hubbard}
In the case of the preparing the Hubbard model ground state, we can verify the electron number parity symmetry
\begin{align*}
S_{\sigma} = \prod_i  \left(1 - 2n_i\right)
\end{align*}
of the output state. Here $n_i$ is the number operator of the $i$th orbital. In the Jordan-Wigner encoding, it simply maps to the product of all $Z$ operators:
\begin{align*}
S_{\sigma} = \prod_i Z_i.
\end{align*}
As shown in \cref{sect:dot_layout}, in the silicon platform, we can measure both the energy terms and the symmetry $S_{\sigma}$ in the same circuit runs, enabling us to perform direct verification. 

In fact, since our ansatz circuit also conserves the electron number within each spin subspace, by using a starting state with the right number of spin-up and spin-down electrons, we can actually verify the electron number parity symmetry within each spin subspace, which we will denote as $S_{\uparrow}$ and $S_{\downarrow}$. 

The ansatz circuit can be decomposed into two-qubit components that represent different interaction terms, fSWAPs and Given rotations as listed in \cref{sect:gate_count_silicon}. We will assume there are $M$ of these components in the circuit, all are affected by depolarising channels with error probability $p$, which means that the mean circuit error count is:
\begin{align}\label{eqn:mean_circ_err}
\mu = Mp.
\end{align}
In between the two-qubit components where the errors occur, our state will have well-defined $S_{\uparrow}$ and $S_{\downarrow}$ (though the set of qubits that correspond to each spin subspace may be different from the final state due to the use of fSWAP). Hence, if a Pauli error occurs here and anti-commutes with $S_{\uparrow}$, it will flip the $S_{\uparrow}$ eigenvalue, leading to a failed  $S_{\uparrow}$ test at the end. Since $S_{\uparrow}$ is the product of $Z$ operators in the spin-up subspace, detectable errors will be the Pauli errors whose total weights in $X$ and $Y$ are odd. Similar arguments also apply to $S_{\downarrow}$ in the spin-down subspace.

There are two types of two-qubit components in the circuit, for which we will make two different approximations to their error channels:
\begin{itemize}
    \item The components acting within \emph{one} spin subspace: Within the $16$ two-qubit Pauli error components (including the identity), we can detect the errors with \emph{one} $X$ or $Y$ in it, which is half of them. Hence, for these two-qubit components, their depolarising errors can be approximated by the composition of a detectable error channel and an undetectable error channel, both with the error probability $\frac{p}{2}$. 
    \item The components acting across \emph{both} spin subspaces: Within the $16$ two-qubit Pauli error components (including the identity), $4$ are undetectable, $4$ are detectable only by $S_{\uparrow}$, $4$ are detectable only by $S_{\downarrow}$ and $4$ are detectable by both. For simplicity, we will absorb the last case into the other two detectable cases and approximate the depolarising channel by the composition of three error channels:
    \begin{enumerate}
        \item An undetectable error channel with the error probability $\frac{p}{4}$.
        \item An error channel detectable by the $S_{\uparrow}$ with the error probability $\frac{3p}{8}$.
        \item An error channel detectable by the $S_{\downarrow}$ with the error probability $\frac{3p}{8}$.
    \end{enumerate}
\end{itemize}
Our circuit consists of alternating layers of these two types of two-qubit components, thus there are approximately equal numbers of components for each type (i.e. $\frac{M}{2}$). Hence, when we focus only on the spin-up subspace, there are around $\frac{M}{4}$ components acting within the subspace with detectable error probability $\frac{p}{2}$ and around $\frac{M}{2}$ components acting across both spin subspaces with detectable error probability $\frac{3p}{8}$, which gives a total detectable mean circuit error count of
\begin{align}\label{eqn:mu_d}
\mu_{d} = \frac{M}{4} \frac{p}{2} + \frac{M}{2} \frac{3p}{8} = \frac{5Mp}{16} = \frac{5\mu}{16}
\end{align}
In the large circuit limit, the probability that $l$ detectable errors occur in the spin-up subspace will follow the Poisson distribution with the mean $\mu_{d}$:
\begin{align*}
P_{l} = e^{- \mu_{d}} \frac{\mu_{d}^l}{l!}.
\end{align*}
The exact same arguments can be applied to the spin-down subspace which gives the same equations. 

When focusing on one of the symmetries, only odd numbers of detectable errors will be detected since an even number of occurrences will commute with $S_\uparrow$/$S_\downarrow$, hence the fraction of circuit runs that pass the verification of $S_\uparrow$/$S_\downarrow$ is:
\begin{align*}
P_{S, one} = \sum_{\text{even } l} P_{l}  = e^{- \mu_d} \cosh(\mu_d) = \frac{1 + e^{- 2\mu_d}}{2}.
\end{align*}
Since in our error channel approximations, events of $S_{\uparrow}$ violation and $S_{\downarrow}$ violation are independent of each other, we can obtain the fraction of circuit runs that pass the verification of both $S_{\uparrow}$ and $S_{\downarrow}$ as:
\begin{align*}
P_S = P_{S, one}^2 = \frac{1}{4} \left(1 + e^{- 2\mu_d}\right)^2.
\end{align*}
Hence, using \cref{eqn:sym_cost} and \cref{eqn:mu_d}, the sampling cost factor of applying symmetry verification using $S_{\uparrow}$ and $S_{\downarrow}$ is
\begin{align}\label{eqn:sym_cost_hubbard}
C_S \sim \frac{1}{P_S} = 4\left(1 + e^{- 2\mu_d}\right)^{-2} = 4\left(1 + e^{- \frac{5\mu}{8}}\right)^{-2}.
\end{align}

When focusing on one of the symmetry, the expected number of circuit errors detectable by $S_{\uparrow}$/$S_{\downarrow}$ after symmetry verification is:
\begin{align*}
\mu_{S, one} = \sum_{\text{even } l} \frac{P_{l}}{P_{S, one}} l =  \frac{\mu_{d}}{P_{S, one}}\sum_{\text{odd } l} P_{l} = \mu_{d} \tanh(\mu_{d}).
\end{align*}
Since the error events of $S_{\uparrow}$ violation and $S_{\downarrow}$ violation are independent of each other, we can obtain the resultant mean circuit error rate after applying symmetry verification using both $S_{\uparrow}$ and $S_{\downarrow}$:
\begin{align}\label{eqn:mu_S}
\mu_S = \mu - 2 \mu_{d} + 2\mu_{S, one} = \frac{3\mu}{8} + \frac{5\mu}{8} \tanh(\frac{5\mu}{16})
\end{align}
where we have made use of \cref{eqn:mu_d}. Hence, the mean circuit error count is reduced by a fraction of:
\begin{align}\label{eqn:sym_err_reduction}
\frac{\mu - \mu_S}{\mu} = \frac{5}{8} \left(1 - \tanh(\frac{5\mu}{16})\right)
\end{align}
after symmetry verification.

Note that the key quantity in our error mitigation analysis is the mean circuit error count $\mu = Mp$, not the gate error rate $p$. Hence, even though we are discussing in terms of two-qubit components here instead of elementary gates, as long as we are using a realistic $\mu$, all of our estimates will be very similar to what we can obtain by considering elementary gates. 

As shown in \cref{eqn:circ_err_rate} and later in \cref{eqn:circ_error_rate_super}, we are interested in the cases in which the original mean circuit error count is around $2$:
\begin{align*}
\mu \sim 2.
\end{align*}
Substituting into \cref{eqn:sym_cost_hubbard}, we can obtain the symmetry verification cost factor:
\begin{align}
C_S \approx 2.4.
\end{align}
After applying symmetry verification using both $S_{\uparrow}$ and $S_{\downarrow}$, we should expect the mean circuit error count reduce by around $30 \%$ using \cref{eqn:sym_err_reduction}.

\subsection{Error Extrapolation}\label{sect:extrapolation}
In the previous section, we have argued that we can reduce the mean circuit error count by a sizeable fraction using symmetry verification. However, the mean circuit error count is still not negligible, thus we need to employ another error mitigation technique -- error extrapolation -- to extract useful information out of our noisy circuit. 

We will write the average value of the observable $O$ obtained from a quantum circuit with mean circuit error count $\mu$ as $\overline{O}_{\mu}$. If we run the circuit with the minimal error rate $\mu$ to obtain $\overline{O}_{\mu}$, and we boost the error rate by a factor of $\lambda$ to obtain $\overline{O}_{\lambda \mu}$, then an estimate of the error-free observable expectation value $\overline{O}_0$ can be obtained via linear extrapolation~\cite{liEfficientVariationalQuantum2017, temmeErrorMitigationShortDepth2017}:
\begin{align}\label{eqn:lin_extrapolate}
\overline{O}_{0} = \frac{\lambda \overline{O}_{\mu} - \overline{O}_{\lambda \mu}}{\lambda - 1}.
\end{align}

Linear extrapolation has been successfully implemented experimentally on superconducting platforms~\cite{kandalaErrorMitigationExtends2019}, in which the error is boosted via changing the time length of the control pulses. Similar methods can be applied to the silicon platform. When the mean circuit error count increase, we can probe at more error rates and go beyond linear extrapolation to Richardson extrapolation to improve our estimation accuracy~\cite{temmeErrorMitigationShortDepth2017}.

For large quantum circuits, there has been numerical evidence showing that it might be more appropriate to extrapolate along an exponential curve instead~\cite{endoPracticalQuantumError2018}. However, the act of symmetry verification might violate some assumptions for exponential extrapolation, thus changing the shape of the extrapolation curve. Hence, in order to combine symmetry verification with error extrapolation, here we will focus on linear extrapolation.

\subsection{Combination}
As shown in \cref{sec:sym_ver_hubbard}, symmetry verification can reduce the mean circuit error count. To combine symmetry verification with error extrapolation, we simply apply symmetry verification to the two observables in \cref{eqn:lin_extrapolate}:
\begin{align}\label{eqn:SE}
\overline{O}_{SE} = \frac{\lambda \overline{O}_{S, \mu} - \overline{O}_{S, \lambda\mu}}{\lambda - 1}.
\end{align}
here $O_{SE}$ is the error-mitigated observable after both symmetry verification and error extrapolation while $O_{S, \mu}$ is the symmetry-verified observable at mean circuit error count $\mu$.

Using \cref{eqn:sym_cost_hubbard}, the cost of symmetry verification is given by:
\begin{align}\label{eqn:c_s_lambda}
C_{S, \lambda} = 4\left(1 + e^{- \frac{5\lambda\mu}{8}}\right)^{-2}.
\end{align}
Hence, using \cref{eqn:SE} and assuming $\var{O_{\mu}} = \var{O_{\lambda \mu}}$, we have
\begin{align*}
\var{O_{SE}} = \frac{\lambda^2C_{S,1} + C_{S, \lambda}}{\left(\lambda - 1\right)^2} \var{O_{\mu}}
\end{align*}
Hence, for each sample of $O_{SE}$, we need to take $C_{SE}$ total samples of $O_{\mu}$ and $O_{\lambda \mu}$ to achieve same precision level, where
\begin{align*}
C_{SE}&= 2\frac{\lambda^2C_{S,1} + C_{S, \lambda}}{\left(\lambda - 1\right)^2}\\
& = 8\frac{\lambda^2\left(1 + e^{- \frac{5\mu}{8}}\right)^{-2} + \left(1 + e^{- \frac{5\lambda\mu}{8}}\right)^{-2}}{\left(\lambda - 1\right)^2}.
\end{align*}
The factor of $2$ is to account for the samples of both $O_{\mu}$ and $O_{\lambda \mu}$. 

For $\mu = 2$, $\lambda = 2$, we have:
\begin{align}\label{eqn:comb_cost}
C_{SE} \sim 25.
\end{align}
i.e. if we want to apply symmetry verification and two-point linear extrapolation with $\lambda = 2$ to our example, we need $25$ times more circuit runs to obtain an estimated observable than sampling from the circuit directly. The combination of symmetry verification and error mitigation was proven to be very effective in numerical simulations~\cite{mcardleErrorMitigatedDigitalQuantum2019}. In our example, we managed to suppress the original mean circuit error count by almost $30\%$ using symmetry verification. Together with the robustness of the variational circuits against local systematic errors, they should enable efficient applications of error extrapolation, and thus allowing us to extract meaningful results out of our noisy quantum circuit. 

\section{VQE Implementation}\label{sect:runtime_est}
\subsection{Parametrising the Hamiltonian Ansatz}\label{sect:parametrisation}
We have outlined the structure of the ansatz circuit in \cref{sect:hubbard_ansatz}, now we will turn to the way we parametrise the gates in the circuit. In the simplest scheme, we can assign a different parameter to each parametrised gate to allow an unconstrained optimisation of the ansatz circuit. However, a large number of parameters will mean a large number of dimensions in the parameter space. This can lead to difficulties in optimisation and may lead to long runtime since we need to probe more directions to obtain the gradient vector. Hence, here we will try to reduce the number of parameters by using the symmetry of the site layout.

For the open-boundary Hubbard model, the site layout has mirror symmetries along the horizontal and vertical direction. Thus the 2D Hubbard grid can be sliced into four equivalent partitions: $N_{eq} = 4$. On top of that if we have the same number of rows and columns, then the site layout also has an additional diagonal mirror symmetry, which gives $N_{eq} = 8$. The ground state of the Hubbard model is expected to follow the same symmetry. 

The input Slater determinant should follows the same layout symmetry, hence so does the ansatz parametrisation. That is, the parametrised gates that represent the corresponding interaction terms in different equivalent partitions, which can be mapped to each other via layout-symmetry transformations, will share the same parameters. 

Hence, the number of parameters in our ansatz is:
\begin{align}\label{eqn:n_para}
N_{para} \approx \frac{N_{para}^{site}V}{N_{eq}} N_{blk}.
\end{align}
where $N_{para}^{site}$ is the number of parameters per site.

Ignoring the boundary case, there will be $5$ interaction terms associated with each site: the repulsion term and the horizontal and vertical hopping terms of the two spins. The input Slater determinant and the output ground state must have the same spin symmetry since our ansatz preserves spins. Hence, the ansatz parametrisation will also have the same spin symmetry. Without spin-flip symmetry, the parameters for the spin-up and spin-down hopping terms will be different, thus $5$ interaction terms means $5$ parameters: $N_{para}^{site} = 5$. With spin-flip symmetry, the gates for the spin-up and spin-down hopping terms can share the same set of parameters, thus $N_{para}^{site} = 3$. If we are considering the half-filling ground state with the smallest total spin, then:
\begin{itemize}
    \item Odd number of sites (electrons): different numbers of spin-up and spin-down electrons, which means no spin-flip symmetry and $N_{para}^{site} = 5$.
    \item Even number of sites (electrons): same number of spin-up and spin-down electrons, which means spin-flip symmetry and $N_{para}^{site} = 3$.
\end{itemize}

For the Hubbard model with periodic boundaries along the horizontal direction, we have translational symmetry of the sites along the horizontal direction, which means that all columns are equivalent on top of the vertical mirror symmetry we have, giving $N_{eq} = 2N_{col}$. This can lead to fewer parameters. However, as we will discuss in \cref{sect:energy_measurement}, the vertical interaction terms cannot be measured locally and efficiently in this case, thus we will not consider such a boundary condition. 

For the Hubbard model with periodic boundaries in both directions, we have complete translational symmetry and thus every site is equivalent. However, we will also not consider this case since there is not yet an efficient Hamiltonian ansatz circuit for the periodic Hubbard model with a gate count and scaling as favourable as the one we have adopted for the open-boundary case.

\subsection{Measurement}\label{sect:energy_measurement}
The measurement of energy is carried out by measuring the individual Pauli components of the Hamiltonian. The Pauli components within an individual repulsion term or a hopping term commute with each other. Thus with the availability of non-demolishing measurements, ideally we should be able to measure all the commuting interaction terms at one go. For the Hubbard model, there are five commuting subsets as shown in \cref{fig:site_order}: repulsion terms, even horizontal hopping terms, odd horizontal hopping terms, even vertical hopping terms and odd vertical hopping terms. Thus we should be able to obtain one sample of each interaction term in five circuit runs. However, direct measurements of the vertical hopping terms can be costly due to their non-locality. These non-local terms can be broken down into non-local Pauli observables, which in turn can be obtained by performing local Pauli measurements and multiplying the results. However, such local measurements can break the commutativity of the vertical hopping terms such that we cannot measure them in parallel.

For the case of the open-boundary Hubbard model, we can tackle this by switching the canonical ordering of orbitals from running horizontally in the 2D grid to running vertically when we try to measure the vertical hopping terms. In such a way, just like the horizontal hopping terms are local in the horizontal-running canonical order, the vertical hopping terms will also be local in the vertical-running canonical order. Note that switching the orbital canonical order will require us to switch the ansatz accordingly, but the same parameters will be used for the parametrised gates that correspond to the same interaction terms. In such a way, we can still obtain one sample of all interactions terms in five circuit runs. 
The same measurement scheme can be used to measure the energy gradients since it involves the same Pauli measurements with some small modifications to the ansatz circuit (see \cref{sect:circuit_grad_meas}). 

Now on top of obtaining the energy or energy gradient by measuring the Pauli components $G_j$ of the Hubbard Hamiltonian, we also want to apply direct symmetry verification outlined in \cref{sect:sym_ver} by measuring the symmetry operator $S$ in the same run. However, it is often the case that $S$ is not local, thus to obtain $S$ we need to rely on local measurements and post-processing.

In our example, there is no need to measure the electron number parity symmetry, it can be obtained by composing the measurement results of the interaction terms. When measuring the repulsion terms, we will be performing single-qubit $Z$ measurement for every qubit. The repulsion terms and the electron number parity can both be obtained via post-processing. In the case of measuring the hopping terms, we will be measuring $XX$ and $YY$ for the hopping pairs adjacent to each other in the canonical order. We can obtain the results of $ZZ$ measurements by composing $XX$ and $YY$ (with an additional $-$ sign), composing with the $Z$ measurements of the qubits not included in the hopping pairs, we can then obtain the electron number parity via post-processing. 

It is worth noting that the Hubbard model simulation is quite friendly for efficient local measurements of the Hamiltonian terms. For more general problems, one might need to turn to more sophisticated measurement schemes~\cite{hugginsEfficientNoiseResilient2019, crawfordEfficientQuantumMeasurement2019, gokhaleMeasurementCostVariational2019}. 

In the above scheme, we have assumed we can carry out non-demolishing measurements, which can be carried out in silicon using ancilla qubits. In \cref{sect:dot_layout}, we outline a possible quantum dot layout that enables an efficient implementation of our measurement scheme.

\subsection{Optimisation Method}\label{sect:optimisation}
We need to employ classical optimisation algorithms to obtain the optimal set of parameters for our parametrised quantum circuit. There are two general approaches, direct search and gradient-based. Direct search involves evaluating the cost function at different points and choosing the next set of points to evaluate based on the known points of the cost function, while gradient-based methods make use of the gradient of the cost function. In our case, since we are searching for the ground state, the cost function that we want to minimise is the energy of the state produced by our quantum circuit.

The energy can be straightforwardly evaluated by measuring the Pauli components of the Hamiltonian as mentioned in \cref{sect:energy_measurement}. As discussed in Ref.~\cite{corbozCompetingStatesModel2014}, to compete with the best classical algorithm for Hubbard model simulation, we need to estimate the energy per site to the precision $\epsilon_{E, site} = 10^{-3} t$. In \cref{sect:grad_n_sample}, we have translated this precision requirement into the precision requirement on the estimates of each Hamiltonian Pauli term. For the $5 \times 5$ Hubbard model, the number of circuit runs needed to estimate the energy to the required precision is:
\begin{align}\label{eqn:nruns_E}
M_{E} \approx 4 \times 10^{5}.
\end{align}

The gradient vector can be evaluated using finite difference, which involves evaluating the energy at two neighbouring points. In the simplest gradient descent scheme, by evaluating the energy points used in finite difference to the precision $\epsilon_{E, site}$ and also choosing the terminating threshold of the change in energy to be $\epsilon_{E, site}$, we will be able to find the local energy minimum with the required precision $\epsilon_{E, site}$ given the right gradient-descent step size. In such a case, we will need twice the number of measurements compared to energy estimation to evaluate the gradient in one direction (since we need to evaluate two energy points). We have $N_{para}$ directions to probe, thus the number of circuit runs needed to evaluate the full energy gradient vector using finite difference is:
\begin{align}\label{eqn:nruns_df_grad}
M_{grad}^{fd} = 2 N_{para} M_{E}.
\end{align}
Using \cref{eqn:n_para} and assuming $N_{blk} \sim V$, for $5 \times 5$ Hubbard model, the number of circuit runs needed to estimate the full gradient vector using finite difference is:
\begin{align*}
M_{grad}^{fd} = 2 N_{para} M_{E} \approx 3.1 \times 10^{8}.
\end{align*}
The precision of the gradient vector obtained here is dependent on the finite difference step size we choose. In \cref{sect:grad_n_sample}, we have outlined how to obtain the optimal step size and the gradient precision $\epsilon_{grad}$ that we can achieve using this optimal step size.

The gradient vector can also be evaluated using direct measurements of a modified ansatz circuit. The two approaches to obtain the gradient were compared in Refs.~\cite{guerreschiPracticalOptimizationHybrid2017, romeroStrategiesQuantumComputing2018}, in which they explain that more measurements are needed in finite difference to overcome the finite step size approximation it makes. We further compare them for our implementation in \cref{sect:grad_n_sample}, taking into account of the fact that we have the many parametrised gates share the same parameters due to the symmetries in the site layout. We found that direct measurements require fewer samples compared to finite difference as the number of gates with shared parameters increases. In \cref{sect:grad_n_sample}, we explain that to achieve the same gradient precision $\epsilon_{grad}$ achieved above using finite difference (with the optimal step size), the number of circuit runs required using direct measurements is:
\begin{align}\label{eqn:nruns_grad}
M_{grad} \approx 2.5 \times 10^7
\end{align}
which is an order of magnitude better than finite difference in this case. 

Direct search methods are generally more effective in noisy and non-smooth problems while for gradient-based methods, the number of function calls needed usually scales better in higher dimensional problems~\cite{koldaOptimizationDirectSearch2003}. We have also proven above that evaluating the full gradient vector is usually much more costly than evaluating an energy point in our implementation. We can see that neither of the approaches are clearly preferred. Various direct search optimisation methods like Nelder-Mead simplex have been successfully implemented experimentally~\cite{peruzzoVariationalEigenvalueSolver2014, shenQuantumImplementationUnitary2017, santagatiWitnessingEigenstatesQuantum2018, collessComputationMolecularSpectra2018, hempelQuantumChemistryCalculations2018,sagastizabalExperimentalErrorMitigation2019,kokailSelfverifyingVariationalQuantum2019, kokailSelfverifyingVariationalQuantum2019} for small-size problems due to the robustness of direct search against noise, while gradient-based method like SPSA, which requires a smaller number of samples than simple gradient descent due to its stochastic nature, have also found success in the simulations of small molecules~\cite{kandalaHardwareefficientVariationalQuantum2017, kandalaErrorMitigationExtends2019, ganzhornGateEfficientSimulationMolecular2019}. With improvements in the quantum hardware noise rate, we will expect gradient-based methods to play a more and more important role in the experimental realisation of VQE, especially considering the success of advanced gradient-based methods like Adam and Adagrad in high dimension noisy optimisation problems in classical machine learning~\cite{goodfellowDeepLearning2017, ruderOverviewGradientDescent2017}. There are also investigations into using machine learning for optimisation~\cite{verdonLearningLearnQuantum2019, wilsonOptimizingQuantumHeuristics2019}, which might have faster convergence rate and higher robustness to noise. In the end, the optimisation scheme is likely to involve a combination of various methods, with the aid of techniques like block-by-block optimisation~\cite{weckerProgressPracticalQuantum2015} and sequential optimisation~\cite{nakanishiSequentialMinimalOptimization2019}. 

Cade~\textit{et al.}~\cite{cadeStrategiesSolvingFermiHubbard2019} have performed numerical simulation of the Hubbard VQE using SPSA optimisation. They followed a three-stage protocol with coarser gradient precision at the beginning and moving to finer and finer gradient precision as the optimisation progress. In the end they evaluated the gradient using finite-difference with the number of circuit runs for each energy estimation being $M_{E} \sim 5 \times 10 ^4$ (this is different from our result above due to different precision requirements). Their $3 \times 3$ Hubbard simulations converge when the number of circuit runs is around $M_{tot} \sim 2 \times 10^8$. There are $30$ parameters in their circuit, using \cref{eqn:nruns_df_grad}, we can translate the result of their simulation into the language of simple gradient descent, the number of `effective' gradient descent iteration in their simulation is just:
\begin{align*}
n_{iter} = \frac{M_{tot}}{M_{grad}^{fd}} = \frac{M_{tot}}{2N_{para} M_E} = 67.
\end{align*}
As mentioned above, the gradient-based method has been shown to scale extremely well with increased problem dimensions in practice. Thus, we can expect the number of `effective' gradient descent iteration needed for $5 \times 5$ Hubbard simulation will also be in the region of 
\begin{align}\label{eqn:n_iter}
n_{iter} \sim 100.
\end{align}

\subsection{Algorithm Runtime for Silicon Spin Qubits}
Here we will estimate the algorithm runtime needed for running the VQE for the $5 \times 5$ open-boundary Fermi-Hubbard model.

From \cref{sect:ansatz_circuit} and \ref{sect:state_prep}, we know the runtime $T_{circ}$ needed for the ansatz circuit with Slater determinant preparation is:
\begin{align*}
T_{circ} &\approx \left(49 + 45 N_{blk}\right)\tau_{1q} + \left(196 + 80 N_{blk}\right)\tau_{2q} + \tau_{in} + \tau_{m}
\end{align*}
where $\tau_{1q}$ and $\tau_{2q}$ are the typical time needed to perform a $\frac{\pi}{2}$ rotation for $Z$ rotation and partial swap respectively, and $\tau_{in}$, $\tau_{m}$ are the time required for qubit initialisation and measurements. 

In silicon quantum dot spin qubits, the $Z$ rotations can be implemented using the Stark shift at the speed $\tau_{1q} \sim 0.1 \mu s$~\cite{hwangImpactFactorsValleys2017}. 
Partial swaps can be implemented using exchange interaction at sub-$ns$ scale~\cite{pettaCoherentManipulationCoupled2005}. Here instead we will assume a two-qubit-gate time of tens of $ns$ to prevent the gate fidelity being limited by the finite voltage rise time \cite{nowackSingleShotCorrelationsTwoQubit2011}.
For readout, a scheme that can achieve more than $98\%$ fidelity in under $6 \mu s$ has been demonstrated~\cite{zhengRapidGatebasedSpin2019}, and a sub-$\mu s$ scheme with $99.7\%$ fidelity has been proposed~\cite{schaalFastGateBasedReadout2020}. For initialisation, the simplest way is via spin relaxation, which will be on the $ms$ timescale. Faster initialisation can be achieved via spin-selective tunnelling from charge reservoirs~\cite{elzermanSingleshotReadoutIndividual2004} or electron shuttling and `hotspot reset'~\cite{srinivasaSimultaneousSpinChargeRelaxation2013, bertrandQuantumManipulationTwoElectron2015, fogartyIntegratedSiliconQubit2018, yangOperationSiliconQuantum2020}. Initialisations at the $\mu s$ scale have been achieved in silicon donor qubits~\cite{morelloSingleshotReadoutElectron2010} and other semiconductor quantum dot qubits~\cite{shulmanSuppressingQubitDephasing2014}. Thus here we will assume the time needed for initialisation plus readout can be reduced to below $100 \mu s$. Hence, for the $5 \times 5$ Hubbard model with $N_{blk} = V = 25$, the runtime needed for each circuit run is around
\begin{align*}
T_{circ} \sim 250 \mu s.
\end{align*}

In \cref{sect:optimisation}, we have obtained the number of circuit runs needed for estimating the energy and the energy gradient vector. However, as mentioned in \cref{sect:error_mitigation}, due to the high mean circuit error count, we need to apply error mitigation. The sampling cost of applying both direct symmetry verification and linear error extrapolation is $C_{SE} \sim 25$ as shown in \cref{eqn:comb_cost}. Thus the number of circuit runs needed to estimate the energy and the energy gradient vector with error mitigation is:
\begin{equation}\label{eqn:N_samples}
    \begin{aligned}
    M_{E}^* &= C_{SE}M_{E} \approx 1 \times 10^{7}\\
    M_{grad}^* &= C_{SE}M_{grad} \approx 6 \times 10^{8}.
    \end{aligned}
\end{equation}
Thus the time needed to evaluate the error-mitigated energy and energy gradient is:
\begin{align*}
T_E & = T_{circ}M_{E}^* =  2500 \text{ s}\\
T_{grad} & = T_{circ}M_{grad}^* = 1.5 \times 10^5 \text{ s} \approx 1.7 \text{ days}.
\end{align*}

Using the simplest optimisation scheme, gradient descent, each iteration step then involves the evaluation of one gradient vector, which requires $1.7$ days. Such a long duration per iteration is hardly practical. However, the time cost is mostly due to the large number of samples needed, thus can be easily solved by running many circuits in parallel in multiple quantum processors. With $200$ processors, the time required for each gradient-descent iteration is reduced to around $10$ minutes, thus making gradient descent feasible runtime-wise even if we require thousands of iterations for convergence. In \cref{eqn:n_iter}, we have estimated the number of iterations required to be on the order of a hundred, which will corresponds to a total runtime of around $1$ day. $200$ processors will mean $10^4$ qubits, which can easily fit onto a single silicon chip along with the classical controls and measurement devices required, giving us one single integrated multi-core processor for the task. We stress that these cores would be independent of one another. Of course, we do not assert that simple gradient descent would necessarily be able to find the solution for the problem size we are considering. However, its runtime feasibility should be indicative of the runtime of the other more advanced gradient-based methods.

\section{Resource Estimates for Superconducting Qubits}\label{sect:resource_est_super}
When we switch from the silicon spin qubits to superconducting qubits, most of our arguments apply except we need to decompose our circuit using a different gate set and a different set of hardware operation times. For the superconducting qubits, a natural two-qubit gate to use for the Hubbard model simulation will be the partial iSWAP  gate~\cite{mckayUniversalGateFixedFrequency2016}, which is implemented using XY-interaction and is just the hopping interaction gate we need to implement. The differences between partial iSWAP and partial SWAP have been discussed by Schuch~\textit{et al.}~\cite{schuchNaturalTwoqubitGate2003}. By using partial iSWAP as our only elementary two-qubit gate, it also enable us to implement all $Z$ rotations virtually~\cite{mckayEfficientZGatesQuantum2017}. As derived in \cref{sect:gate_count_super}, for the $5 \times 5$ Hubbard model with the number of ansatz blocks equal to the number of sites, the gate counts are:
\begin{align*}
N_{1q} & \approx 2500 \\
N_{2q} & \approx 14000.
\end{align*}
Comparing to \cref{eqn:gate_count_25}, we can see a massive decrease in the single-qubit gate count due to the use of virtual $Z$ gates, which in turn reduces our fidelity requirement for the single-qubit gates. However, applying virtual $Z$ gate would require us to implement partial iSWAP with a range of different frequency tunings~\cite{mckayEfficientZGatesQuantum2017}, which would increase the difficulties in calibrating the partial iSWAP gates. 
We also see a reduction in the number of two-qubit gates required, but it is still of the same order and thus will lead to a similar gate fidelity requirement $ \sim 10^{-4}$. The resultant mean circuit error count is reduced to
\begin{align}\label{eqn:circ_error_rate_super}
\mu \sim 1.5,
\end{align}
which will lead to improved performance of the error mitigation techniques. Individual superconducting quantum processor of size $\sim 50$ has been experimentally demonstrated to be able to perform certain tasks that cannot be performed efficiently on any classical computers~\cite{aruteQuantumSupremacyUsing2019}. There have also been demonstrations of successful implementations of various error mitigation techniques on the superconducting platform~\cite{kandalaErrorMitigationExtends2019, chiesaQuantumHardwareSimulating2019}.

The runtime required for one circuit run as derived in \cref{sect:gate_count_super} is
\begin{align*}
T_{circ} &\approx  125 \tau_{1q} +  650\tau_{2q} + \tau_{in} + \tau_{m}.
\end{align*}
where $\tau_{1q}\, \tau_{2q},\ \tau_{in}$ and $\tau_{m}$ are the time required for single-qubit gates, two-qubit gates (iSWAP), initialisation and readout, respectively. In superconducting qubits, we have $\tau_{1q}\sim 20 ns$~\cite{wendinQuantumInformationProcessing2017}, $\tau_{2q} \sim 200 ns$~\cite{mckayUniversalGateFixedFrequency2016} and $\tau_{in} \sim \tau_{m} \sim 100 ns$~\cite{reedFastResetSuppressing2010, walterRapidHighFidelitySingleShot2017}. Thus the total runtime is around:
\begin{align*}
T_{circ} &\approx 150 \mu s
\end{align*}
which is similar to the runtime estimate for silicon qubits. The difference is that the runtime bottleneck here is the two-qubit gate speed while in silicon the two-qubit gates contribute the least to the overall runtime compared to the other operations. 

Using \cref{eqn:N_samples}, the time needed to evaluate the error-mitigated energy and energy gradient are:
\begin{align*}
T_E & = T_{circ}M_{E}^* =  1500 \text{ s}\\
T_{grad} & = T_{circ}M_{grad}^* = 1 \times 10^5 \text{ s} \approx 1.2 \text{ days}
\end{align*}
which are also similar to the silicon platform. 

Parallelisation in multiple quantum processors is still essential to bring the total runtime down to a practical level. We need around $150$ superconducting-qubit quantum processors, each with $50$ qubits (thus a total of around $7500$ qubits), to bring time required for each gradient-descent iteration to around $10$ minutes. The number of quantum processors required is similar to that of the silicon spin qubits. However, unlike silicon spin qubits which can easily fit all of the required quantum processors in the same chip, the superconducting processors likely need to be distributed into multiple chips due to a lower qubit density.

\section{Conclusion}\label{sect:conclusion}
In this Article, we have investigated the resource requirements on obtaining the ground state of a Hamiltonian in a quantum computer using VQE, in which the Hamiltonian cannot be solved exactly classically. The Hamiltonian we have chosen is the $5 \times 5$ open-boundary Fermi-Hubbard model due to its favourable scaling in both circuit size and depth. We began our analysis by considering silicon spin qubits as our example hardware platform for the resource estimation. Our ansatz circuit makes use of one of the latest schemes for the input Slater determinant preparation~\cite{jiangQuantumAlgorithmsSimulate2018} and the Hubbard Hamiltonian ansatz implementation~\cite{kivlichanQuantumSimulationElectronic2018}, which translates into $17000$ single-qubit gates (all are $Z$ rotations) and $26000$ two-qubit gates (all are partial swaps) assuming the number of Hamiltonian blocks in the ansatz is equal to the number of sites. Hence, with perfect single-qubit $Z$ rotations and a two-qubit gate error rate on the order of $10^{-4}$, we can achieve a circuit error of $\sim 2.5$. To obtain meaningful results with this mean circuit error count, we must incorporate error mitigation techniques like error extrapolation and symmetry verification for which we have discussed their combined application to our example. We have devised a measurement scheme that allows us to estimate various terms in the Hamiltonian in parallel and apply symmetry verification at the same time. Bringing all these together, we have estimated the runtime needed for one circuit execution on the silicon-spin-qubit platform to be around $250 \mu s$  and thus one iteration in a simple gradient descent optimisation is around $1.7$ days due to the large number of samples needed. Hence, we have to run VQE in parallel across multiple quantum processors to reduce the runtime to a feasible level. Around $200$ quantum processors with $50$ data qubits each, which can easily fit onto a single silicon chip, can reduce the time required for one gradient-descent iteration to around $10$ minutes, and the estimated total algorithm runtime in this case will be around 1 day. 

We then applied the same analysis to superconducting qubits. We found a large reduction in the number of single-qubit gates due to the ability to implement all $Z$ rotation virtually by using partial iSWAP as our elementary two-qubit gate. By assuming a two-qubit gate error rate of $10^{-4}$ and an equal or lower single-qubit gate error rate, we can achieve a mean circuit error count of $\sim 1.5$. To implement the same algorithm, the circuit runtime is around $150 \mu s$ and thus one gradient-descent iteration takes around $1.2$ days. This can be reduced to around $10$ minutes if we parallelise our task across $150$ superconducting quantum processors. Resource estimates for other qubit platforms can be readily obtained by following similar arguments.

To implement Hubbard VQE on NISQ machines, the first key difficulty is to maintain the mean circuit error count at the order of unity or less so that error mitigation can be effective. Superconducting qubits have the advantage here due to the lower gate counts from the good fit of its elementary gate set to the problem, and the recent demonstration of high performance superconducting quantum processors of the relevant size~\cite{aruteQuantumSupremacyUsing2019}. The second key difficulty is the extremely long runtime due to the huge number of circuit runs required, which motivates the necessity of parallelising our tasks over hundreds or thousands of quantum processors. Silicon spin qubits have better potential here due to its higher qubit density, enabling us to easily fit all of these quantum processing units into one single multi-core processor, leaving a smaller and more manageable physical footprint. 

We can see that implementing a 50-qubit Hubbard model VQE on a NISQ machine sits right at the boundary of being practical in terms of gate counts and mean circuit error count. Hence, even a constant factor improvement in the mean circuit error count can have big impacts on bringing such an application of NISQ machines into reality, which can be brought about via further optimisation of our simulation schemes, improvements in the gate fidelity, improvements in the optimisation scheme and advances in the error mitigation techniques. We also need to rely on parallelisation to tackle the runtime issue. It is worth noting that the number of qubits required in noisy VQE can become comparable to the fault-tolerant implementation. In the case of VQE, what we need is a lot of independent small units for parallelisation instead of a single integrated device, which should massively reduce the difficulties in manufacturing and calibration even if the total number of qubits is of the same order. However, this also places great emphasis on the ability of the hardware platform to reproduce many copies of the quantum processor once we manage to manufacture a good one. Both of the platforms we discussed have advantages in this respect since their compatibility with commercial CMOS fabrication technology can provide a high-precision, relatively low-cost and highly reproducible manufacturing process.

Due to the large number of hyper-parameters in VQE, our gate and runtime estimates are only indicative of the canonical case. There can be variability in the estimates when we choose a different set of hyper-parameters. One important assumption we made is the number of ansatz blocks in the circuit is equal to the number of sites: $N_{blk} = V$, which is an optimistic assumption given what we have observed in the numerical simulations for small size system~\cite{weckerProgressPracticalQuantum2015, reinerFindingGroundState2019, cadeStrategiesSolvingFermiHubbard2019}. An increase in $N_{blk}$ will lead to a linear increase in the gate counts and a quadratic increase in the runtime (due to the increase in both the gate counts and the parameter counts). 

Any increase in the gate counts will lead to an increase in the mean circuit error count, which needs to be suppressed using stronger error mitigation. For example, instead of two-point error extrapolation, we can sample at more error rates, which will increase the number of circuit runs. We can also try to verify for additional symmetries. However, when we have multiple symmetries, it might not be possible to measure the energy terms along with all the symmetries using local measurements in a single circuit run any more, which may lead to additional runtime overhead. It is also worth exploring the possible combinations with other error mitigation techniques like quasi-probability~\cite{temmeErrorMitigationShortDepth2017, endoPracticalQuantumError2018}, to try to gain improvements in estimation accuracy and/or sampling costs. We can also try to develop new error mitigation techniques. One possible avenue could be tailoring the noise in real machines using simple gates to increase the sensitivity of our symmetry verification against the noise. The effectiveness of a similar idea in the context of a quantum error correction code has been shown~\cite{caiMitigatingCoherentNoise2020}.

Any increase in the algorithm runtime can be mitigated via further parallelisation by adding more quantum processors. Without any sudden large changes in the parameters during the optimisation, we should expect the energy and the energy gradient change in a relatively smooth manner as the optimisation progresses. Hence, we can use the energy and the energy gradients we obtained in the previous steps as the prior for the estimation of the new energy and the energy gradients in a Bayesian manner~\cite{mccleanHybridQuantumclassicalHierarchy2017}. This should enable us to achieve the same precision using much fewer samples and thus much shorter algorithm runtime. Other important factors that can influence the algorithm runtime include the way we parametrised our circuit and the exact optimisation algorithm we choose, both worth further explorations.

Since our ansatz circuit has a relatively short depth ($\mathcal{O}(N^{\frac{1}{2}})$), the main limiting factor preventing us from simulating the Hubbard model much beyond the size $5 \times 5$ is due to the increased gate counts and the resultant increased mean circuit error count. Stronger error mitigation or even new error mitigation can only alleviate this problem. To fully tackle this, one possibility is to switch to a different kind of qubit encoding: the Majorana loop stabiliser code~\cite{jiangMajoranaLoopStabilizer2019}. This encoding allows local vertical hopping terms using mediator qubits, thus no Fermionic swap network is needed and we can achieve a Hamiltonian ansatz block and a Slater determinant preparation circuit with depth $\mathcal{O}(1)$. Furthermore, it can detect and correct single-qubit errors. Hence, it can potentially suppress circuit errors and take us much beyond our current problem size. However, more qubits are needed for the encoding and we need to implement higher-weight operators for the interaction terms. In addition, when considering the stabiliser checks, we need to consider the errors introduced in implementing the check circuits and the connectivity requirement on the hardware. On the other hand, we may implement these stabiliser checks in a post-processing way similar to symmetry verification~\cite{mccleanDecodingQuantumErrors2020}, but many more circuit runs will be needed.

\section*{Acknowledgements}
The author would like to thank Sam McArdle, Yuan Xiao and Simon Benjamin for reading through the manuscript and providing valuable insights. The author would also like to thank Sofia Patom{\"a}ki, Michael Fogarty and John Morton for discussions on the hardware implementation.

The author acknowledges support from Quantum Motion Technologies Ltd.

\newpage
\appendix
\section{Hubbard Model Hamiltonian Ansatz}\label{sect:ansatz_circuit}
\subsection{Simulation scheme}\label{sect:sim_scheme}
\begin{figure}[htbp]
    \centering
    \subfloat[]{\includegraphics[width = 0.2\textwidth]{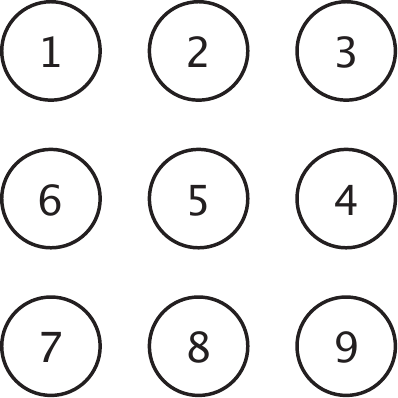}}\\
    \subfloat[]{\includegraphics[width = 0.45\textwidth]{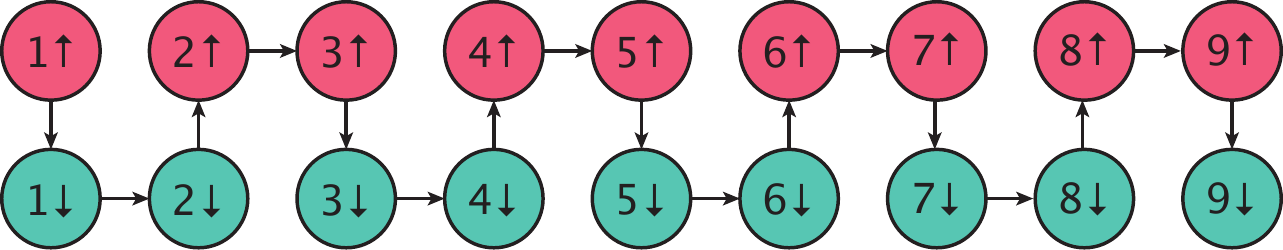}}
    \caption{(a) The layout and labelling of the sites in a 3-by-3 Hubbard model. (b) The corresponding canonical ordering of the orbitals at the beginning of the circuit. The two rows of different colours denote different spins.}
    \label{fig:layout}
\end{figure}

Here we recap the scheme to implement the 2D open-boundary Hubbard model Hamiltonian ansatz as described in \cite{kivlichanQuantumSimulationElectronic2018}. We will be considering a 2D Hubbard models of $V$ sites, with the starting canonical ordering of the orbitals as shown in \cref{fig:layout}.

\begin{figure}[htbp]
    \centering
    \includegraphics[width = 0.45\textwidth]{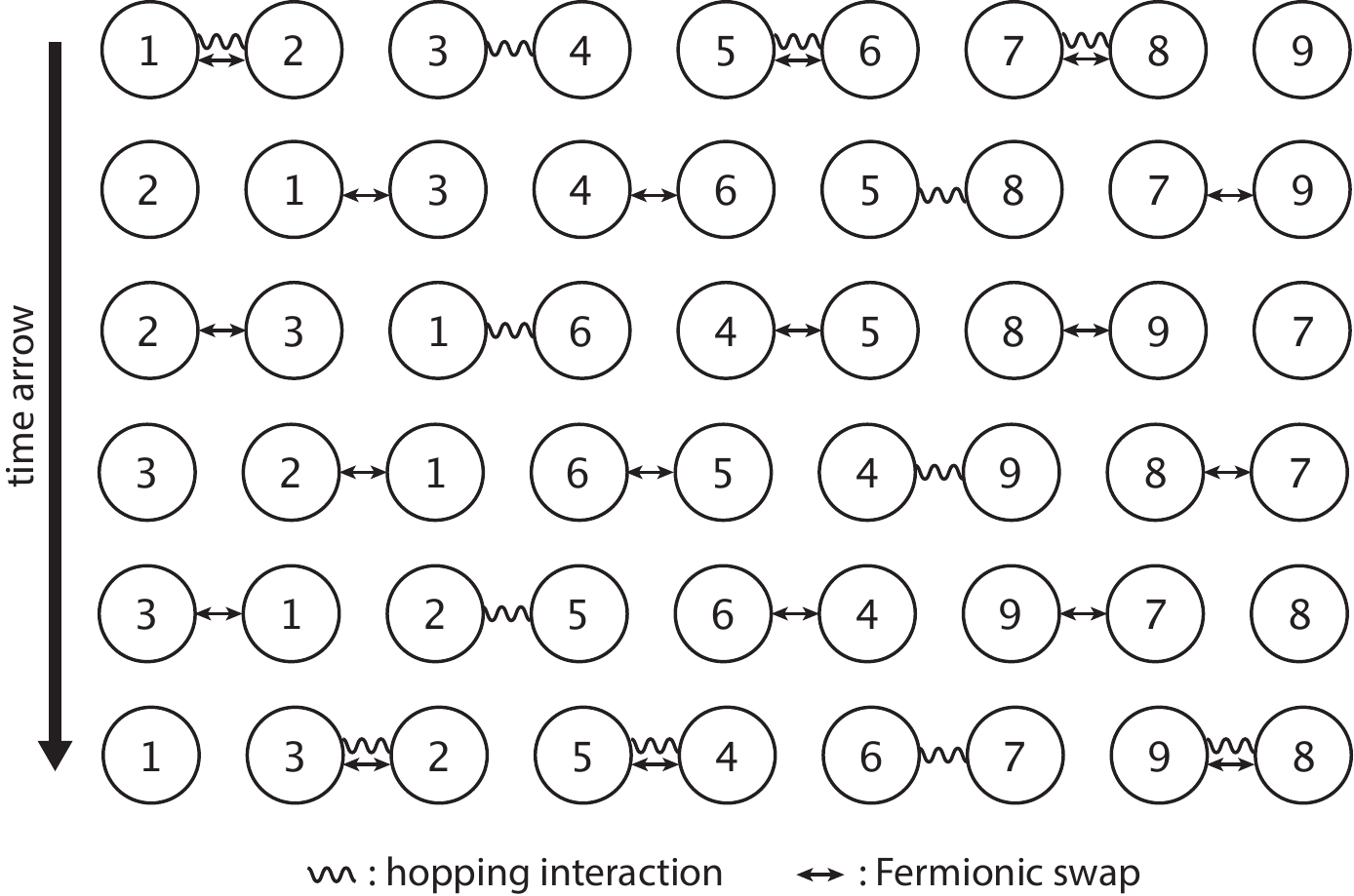}
    \caption{The scheme for swapping spin-down orbitals and applying the relevant hopping interactions for a $3\times 3$ Hubbard model. In between steps, there are also swaps between the two spins that we did not show. In the end, the spin orbitals return to the original ordering, enabling the application of the next block of the Hamiltonian ansatz. A similar scheme is applied to the spin-up orbitals except that it starts with odd-pair swaps first.}
    \label{fig:swap_scheme}
\end{figure}

Gates will be local if they are applying to the orbitals adjacent to each other in the canonical ordering. In the canonical ordering shown in \cref{fig:layout}, the orbitals of the same site and different spins are adjacent to each other, enabling the application of local parametrised on-site repulsion gates. To apply local hopping gates, we need to apply fermionic swaps to the orbitals to move them around in the canonical ordering. There are two types of fermionic swaps that we can apply: swaps between or within spins, which corresponding to swaps between or within rows for the orbital layout in \cref{fig:layout}. The swap scheme in \cite{kivlichanQuantumSimulationElectronic2018} involving alternating swaps within and between spins. Now suppose we focus only on the spin-down orbitals. The spin-down orbitals start in row 2 in which we can only perform local swaps and local hopping interactions between the even pairs of orbitals. Then we swap between spins, moving spin down to row 1, and now we can do local swaps and local hopping interactions between the odd pairs of orbitals. Repeating these steps will enable us to alternate between odd- and even-pair swaps and hopping interactions within the spin-down orbitals, which are interleaved with swaps between the two spins. Such a scheme can apply all relevant hopping interactions in a local manner as shown in \cref{fig:swap_scheme}.

We will define an \emph{edge pair} as a pair of sites that are adjacent to each other in the canonical ordering and are vertical neighbours in the site layout. For example, in \cref{fig:layout}, the edge pairs are $(3,4)$ and $(6,7)$.
The way we perform one block of Hamiltonian ansatz is by repeating $2 * N_{col}$ rounds of the following two steps:
\begin{enumerate}
    \item Swap between spins: swapping orbitals of the same sites but different spins.
    \item Swap within the same spins, except for edge pairs on which we perform hopping interaction instead.
\end{enumerate}
This will return all the orbitals to their original positions. Within the process, we need to 
\begin{itemize}
    \item Substitute step 1 in the first round with repulsion interactions and swaps between spins.
    \item Substitute step 2 in the first and last round with hopping interactions and swaps interaction within same spins, except for edge pairs on which we only perform hopping interactions, no swaps.
\end{itemize}
We can also implement the periodic boundary condition in the horizontal direction by adding a pair of hopping interaction into certain rounds of step 2.

\subsection{Gate count analysis for ansatz}\label{sect:gate_count_silicon}
The gates in the Hubbard model simulation scheme are parametrised gates for on-site repulsion, adjacent-site hopping, fermionic swaps and combinations of them. Here we will decompose them into single-qubit rotations and partial swaps, which form one of the basic gate sets in silicon qubits. 

We will find the hopping gates, the fermionic swaps, and the fSWAP+hopping all starts with $Z_{\frac{\pi}{2}}$ and end with $Z_{-\frac{\pi}{2}}$ on one of the qubits. For on-site repulsion and fSWAP+repulsion, we have $Z_{-\frac{\pi}{2}}$ at the end and we can easily add an $Z_{\frac{\pi}{2}}$ in front by adding a Z rotation pair $Z_{\frac{\pi}{2}}Z_{-\frac{\pi}{2}}$. All of these gates are symmetric under the exchange of qubits, thus we can choose which qubit to place the $Z$ rotations. 

We will choose to place these $Z$ rotation on the odd qubits in the spin-up space and on the even qubits in the spin-down space. In this way, these $Z$ rotation will cancel \footnote{Note that in each iteration, we will have $\sqrt{V} - 1$ interactions within the same spin, while we have $\sqrt{V}$ interaction in between different spins, hence, the $Z$ gate on one of the dot will not be cancelled, this will be the last dot in either the spin up space or the spin-down space. Also, the $Z$ rotations at the beginning and the end of the circuit are not cancelled. However, when estimating the number of gates needed, for a large number of sites (hence large number of rounds of iterations), we will assume such boundary effects are negligible. }. For the hopping term, the fermionic swap and the fSWAP+hopping, this means removing two single-qubit rotations. For on-site repulsion and fSWAP+repulsion in which we have added a Z rotation pair $Z_{\frac{\pi}{2}}Z_{-\frac{\pi}{2}}$ in front, the gate counts do not change.

In the following section, we will use the below notations for the time units of the gates:
\begin{itemize}
    \item $\tau_{1q}$: the time unit for a single-qubit gate, which is the time needed to perform a $\frac{\pi}{2}$ rotation. We will assume the time needed to carry out a single-qubit gate with a variable parameter, i.e. gates like $Z_\theta$, is on average $\tau_{1q}$.
    \item $\tau_{2q}$: the time unit for a two-qubit gate, which is the time needed to perform a $\sqrt{\text{SWAP}}$. We will assume the time needed to carry out a partial swap with a variable parameter, i.e. gates like $\text{SWAP}_\theta$, is on average $\tau_{2q}$.
\end{itemize}
The decompositions of the gates into partial swaps and single-qubit rotations and their resource estimates are shown below:
\begin{itemize}
    \item On-site repulsion: $U_{U}(\theta) = e^{-i \frac{\theta}{2} \left(I-Z_1\right)\left(I-Z_2\right)}$
    \begin{center}
        \begin{tikzcd}
            &\gate[wires=2, style={yshift=-9.5pt, inner ysep=-5pt}]{\rotatebox{270}{$\text{SWAP}_{2\theta}$}}
            &\qw
            &\gate[wires=2, style={yshift=-9.5pt, inner ysep=-5pt}]{\rotatebox{270}{$\text{SWAP}^{\frac{3}{2}}$}} &\gate{Z_{-2\theta}} &\gate[wires=2, style={yshift=-9.5pt, inner ysep=-5pt}]{\rotatebox{270}{$\sqrt{\text{SWAP}}$}}         &\qw&\qw\\
            &&\gate{Z_{\frac{\pi}{2}}}& &\qw& &\gate{Z_{-\frac{\pi}{2}}}&\qw
        \end{tikzcd}
    \end{center}
    Gate counts: $G_{1q, U} = 3$, $G_{2q, U} = 3$.
    
    Time needed: $\tau_U = 4\tau_{1q} + 6\tau_{2q}$
    
    \item Hopping interaction: $U_{t}(\theta) = e^{-i \frac{\theta}{2} (XX + YY)}$
    \begin{center}
        \begin{tikzcd}
            &\qw
            &\gate[wires=2, style={yshift=-9.5pt, inner ysep=-5pt}]{\rotatebox{270}{$\text{SWAP}^{\frac{3}{2}}$}} &\gate{Z_{\theta}} &\gate[wires=2, style={yshift=-9.5pt, inner ysep=-5pt}]{\rotatebox{270}{$\sqrt{\text{SWAP}}$}}         &\qw &\qw\\
            &\gate{Z_{\frac{\pi}{2}}}& &\gate{Z_{-\theta}}& &\gate{Z_{-\frac{\pi}{2}}}&\qw
        \end{tikzcd}
    \end{center}
    Gate counts: $G_{1q, t} = 2$, $G_{2q, t} = 2$
    
    Time needed: $\tau_{t} = \tau_{1q} + 4\tau_{2q}$
    \item Fermionic swap: \\$F_{sw} = \frac{1}{2} \left(XX + YY + ZI + IZ\right) = \text{SWAP}\cdot\text{CZ}$
    \begin{center}
        \begin{tikzcd}
            &\qw
            &\gate[wires=2, style={yshift=-9.5pt, inner ysep=-5pt}]{\rotatebox{270}{$\text{SWAP}^{\frac{3}{2}}$}}    &\gate{Z_\pi} &\gate[wires=2, style={yshift=-9.5pt, inner ysep=-5pt}]{\rotatebox{270}{$\sqrt{\text{SWAP}}$}}         &\qw &\qw\\
            &\gate{Z_{\frac{\pi}{2}}}&     &\qw&  &\gate{Z_{-\frac{\pi}{2}}}&\qw
        \end{tikzcd}
    \end{center}
    Gate counts: $G_{1q, F} = 1$, $G_{2q, F} = 2$
    
    Time needed: $\tau_f = 2\tau_{1q} + 4\tau_{2q}$
    \item Fermionic swap + on-site repulsion: $F_{sw}U_{U}$
    \begin{center}
        \begin{tikzcd}
            &\gate[wires=2, style={yshift=-9.5pt, inner ysep=-5pt}]{\rotatebox{270}{$\text{SWAP}_{2\theta}$}}
            &\qw
            &\gate[wires=2, style={yshift=-9.5pt, inner ysep=-5pt}]{\rotatebox{270}{$\text{SWAP}^{\frac{3}{2}}$}} &\gate{Z_{-2\theta+\pi}} &\gate[wires=2, style={yshift=-9.5pt, inner ysep=-5pt}]{\rotatebox{270}{$\sqrt{\text{SWAP}}$}}         &\qw&\qw\\
            &&\gate{Z_{\frac{\pi}{2}}}& &\qw& &\gate{Z_{-\frac{\pi}{2}}}&\qw
        \end{tikzcd}
    \end{center}
    
    Gate counts: $G_{1q, FU} = 3$, $G_{2q, FU} = 3$.
    
    Time needed: $\tau_{FU}  = 6\tau_{1q} + 6\tau_{2q}$
    \item Fermionic swap + hopping interaction: $F_{sw}U_{t}$
    \begin{center}
        \begin{tikzcd}
            &\qw
            &\gate[wires=2, style={yshift=-9.5pt, inner ysep=-5pt}]{\rotatebox{270}{$\text{SWAP}^{\frac{3}{2}}$}} &\gate{Z_{\theta+ \pi}} &\gate[wires=2, style={yshift=-9.5pt, inner ysep=-5pt}]{\rotatebox{270}{$\sqrt{\text{SWAP}}$}}         &\qw &\qw\\
            &\gate{Z_{\frac{\pi}{2}}}& &\gate{Z_{-\theta}}& &\gate{Z_{-\frac{\pi}{2}}}&\qw
        \end{tikzcd}
    \end{center}
    Gate counts: $G_{1q, Ft} = 2$, $G_{2q, Ft} = 2$
    
    Time needed: $\tau_{Ft} = 3\tau_{1q} + 4\tau_{2q}$
\end{itemize}
Following the above gate decomposition and the Hamiltonian ansatz circuit outlined in \cref{sect:sim_scheme}, we can obtain the following estimates for the total number of one-qubit gates needed $N_{1q}$ (all are $Z$ rotations), total number of two-qubit gates needed $N_{2q}$ (all are partial swaps) and the total length of time needed $T$ to perform one block of Hamiltonian ansatz for 2D Hubbard model of $V$ sites using basic single-qubit rotation and partial swaps:
\begin{align*}
N_{1q} & = 4V^{\frac{3}{2}} + 7V - 4\sqrt{V}\\
N_{2q} & = 8V^{\frac{3}{2}} + V - 4\sqrt{V}\\
T &=\left( 8\sqrt{V}+ 5\right) \tau_{1q} + \left(16\sqrt{V} + 2\right)\tau_{2q}
\end{align*}

For $V = 25$, we have:
\begin{align*}
N_{1q} & \approx 650\\
N_{2q} & \approx 1000\\
T & \approx 45 \tau_{1q} + 80 \tau_{2q}
\end{align*}

\section{Slater Determinant Preparation}\label{sect:state_prep}
Here we recap the Slater determinant preparation scheme outlined in~\cite{jiangQuantumAlgorithmsSimulate2018, kivlichanQuantumSimulationElectronic2018}. Note that the input Slater determinant we choose to prepare in this Article will follow the same spin and site-layout symmetry as the output ground state since the ansatz we choose preserves these two symmetries.
\subsection{Background}
We will use $N_{orb}$ to denote the total number of orbitals that we are considering while $N_{e}$ will be the number of electrons (i.e. the number of occupied orbitals).

We start with the qubits representing the eigenorbitals of the initial Hamiltonian (e.g. non-interacting Hubbard Hamiltonian). The initial state is the ground state of the initial Hamiltonian with the first $N_e$ orbitals being occupied (i.e. the first $N_e$ qubits are initialised to $1$ while the rest are initialised to $0$). Now the role of the state preparation circuit is to transform our qubits from representing the eigenstates of the initial Hamiltonian (with orbital creation operators $\{a_j^\dagger\}$) to the orbitals that can have a compact description of our target Hamiltonian, i.e. to the site orbital basis (with orbital creation operators $\{b_j^\dagger\}$). This basis transformation can be described by the transformation matrix $Q^\dagger$ (also called the Slater determinant) of the shape $N_{orb} \times N_e$:
\begin{align*}
\vec{b}^\dagger &= Q^\dagger \vec{a}^\dagger\\
& = U^\dagger \Lambda W \vec{a}^\dagger\\
& = U^\dagger \Lambda \vec{a}'^\dagger\\
\end{align*}
Here we have carried out singular value decomposition of $Q^\dagger$. $W$ is a rotation within the filled-orbital subspace to find a new set of basis $\{a_i'^\dagger\}$ other than the eigenstate of the initial Hamiltonian $\{a_i^\dagger\}$. The qubit ground state in basis $\{a_i'^\dagger\}$ is the same as basis $\{a_i^\dagger\}$ with all the orbitals filled. Hence, we do not need to carry out the transformation $W$ explicitly, we only need to keep in mind that now we are working in this new basis for the input state instead of the initial basis~\cite{jiangQuantumAlgorithmsSimulate2018}. $\Lambda$ is a rectangular matrix of the $N_{orb} \times N_e$ with ones at the diagonal and zeros elsewhere. This is just an isometry to expand our space from the filled orbital subspace to the full orbital space by attaching $N_{orb} - N_e$ empty orbitals. Then the transformation $U^\dagger$ on the full orbital space will complete our transformation of basis. The transformation $U^\dagger$ will be implemented as compositions of Givens rotations in the quantum circuit.

\subsection{Givens rotation}
A Givens rotation is just a general rotation operation within a 2D complex subspace. There are two parts to a Givens rotation, one is the rotation to change the amplitude, and the other is phase operator to change the relative phase between the two basis in the subspace~\cite{jiangQuantumAlgorithmsSimulate2018}:
\begin{align*}
G(\theta, \phi) = \begin{pmatrix}
\cos \theta & -\sin \theta \\ \sin \theta & \cos \theta
\end{pmatrix}
\begin{pmatrix}
1&0\\0&e^{i\phi}
\end{pmatrix}
\end{align*}
We will be mostly dealing with real Slater determinant, so let us ignore the phase part here. The rotation part can be carried out using the operator:
\begin{align*}
R_{ij}(\theta) &= e^{\frac{\theta}{4}(a_i^\dagger a_j - a_j^\dagger a_i)}
\end{align*}
JW transform the neighbouring orbital case, we have
\begin{align*}
\left(a_i^\dagger a_{i+1} - \left(a_i^\dagger a_{i+1}\right)^\dagger\right)& = 2i\Im{a_i^\dagger a_{i+1}}\\
&\Rightarrow 2i\Im{(X - iY)(X+iY)}\\
& = 2i(XY - YX)
\end{align*}
Hence, the given rotation for adjacent orbitals are:
\begin{align*}
R(\theta) &= e^{i\frac{\theta}{2}(XY - YX)} = e^{-i\frac{\theta}{2}(YX - XY)}
\end{align*}
which translate into the following circuit using partial swap and $Z$ rotations.
\begin{center}
    \begin{tikzcd}
        &\gate[wires=2, style={yshift=-9.5pt, inner ysep=-5pt}]{\rotatebox{270}{$\text{SWAP}^{\frac{3}{2}}$}} &\gate{Z_{-\theta}} &\gate[wires=2, style={yshift=-9.5pt, inner ysep=-5pt}]{\rotatebox{270}{$\sqrt{\text{SWAP}}$}}         &\qw\\
        & &\gate{Z_{\theta}}& &\qw
    \end{tikzcd}
\end{center}
Using the notation in \cref{sect:gate_count_silicon}, we have:
\begin{itemize}
    \item[]Gate counts: $G_{1q, G} = 2$, $G_{2q, G} = 2$\\
    Time needed: $\tau_G = \tau_{1q} + 4\tau_{2q}$
\end{itemize}
\subsection{Gate count for Givens rotation}\label{sect:gate_count_state_prep}
\subsubsection{Simple scheme}
The transformation from the initial eigenbasis $\{a^\dagger_i\}$ to the target eigenbasis $\{b^\dagger_i\}$ can be viewed as a process of trying to diagonalise the transformation matrix (Slater determinant) $Q$. The transformation $W$ can zero out a triangle of entries along the $N_e$ dimension. We need one Givens rotation to zero out each remaining non-zero off-diagonal elements. The number of Givens rotation needed is~\cite{jiangQuantumAlgorithmsSimulate2018}:
\begin{align*}
N_e \left(N_{orb} - N_e\right)
\end{align*}
For the half-filling Hubbard model with $V$ sites, we have $N_{orb} = 2 V$ and $N_e = V$. Hence, we have $V^2$
off-diagonal elements to be zeroed out. Each Givens rotation is decomposed into two partial-swaps and two $Z$ rotations, thus the number of gates we needed are (following the notation in \cref{sect:gate_count_silicon})
\begin{align*}
N_{1q} = 2V^2\\
N_{2q} = 2V^2
\end{align*}

Using the parallel scheme suggested in~\cite{kivlichanQuantumSimulationElectronic2018}, we have a circuit of depth
\begin{align*}
\left(N_{orb} - 1 - \left(N_e - 1\right)\right) + \left(N_e - 1\right)  = N_{orb} - 1 = 2V-1
\end{align*}
which translate into the circuit runtime of
\begin{align*}
T = \left(2V-1\right)\left(\tau_{1q} + 4\tau_{2q}\right)
\end{align*}

For $V = 25$, we have:
\begin{align*}
N_{1q} &= 1250\\
N_{2q} &= 1250\\
T &= 49 \tau_{1q} + 196 \tau_{2q}
\end{align*}

\subsubsection{Using spin conservation}
For the Hubbard Hamiltonian that we are considering, the two spin spaces are decoupled. Hence, we can consider the Slater determinants of the two spin subspace separately, each of the shape $\frac{N_e}{2} \times \frac{N_{orb}}{2}$. Hence, the number of Givens rotation needed in each spin subspace is
\begin{align*}
\frac{N_e}{2} \left(\frac{N_{orb}}{2} - \frac{N_e}{2}\right) = \frac{V^2}{4}.
\end{align*}
To keep the applications of the Givens rotations within the spin subspace on only adjacent orbitals, we need to start in an orbital ordering with the first $N_{orb}/2$ being the spin-up orbitals and the next $N_{orb}/2$ being the spin-down orbitals. Hence, we need to carry out orbital rearrangement after the Givens rotations to restore the spin-orbital ordering for the Hamiltonian ansatz (\cref{sect:sim_scheme}).

We can arrange the Givens rotations such that all the Givens rotation on the last spin-up orbital (the $\frac{N_{orb}}{2}$th orbital) and the first spin-down orbital (the $\frac{N_{orb}}{2} + 1$th orbital) finished first. In the next time step, while we are carrying out other Givens rotations, we can start performing fSWAP on the two orbitals that have finished Givens rotation. In the next step, two more orbitals will finish their Givens rotation (orbitals $\frac{N_{orb}}{2} -1$ and $\frac{N_{orb}}{2} + 2$), thus now we can perform fSWAP on orbitals $(\frac{N_{orb}}{2} -1, \frac{N_{orb}}{2})$ and $(\frac{N_{orb}}{2} +1, \frac{N_{orb}}{2}+2)$. Carry on we will have more and more orbitals finishing their Givens rotations, and in each time step we will perform fSWAP on all orbitals that finished Givens rotations, alternating between the odd pair of orbitals and even pair of orbitals. After $\frac{N_{orb}}{2} - 2$ layer of swap, we will have alternating up and down orbitals. One more layer of swaps between all orbitals $4n -1$ and $4n$, will give us the $\uparrow \downarrow \downarrow \uparrow \uparrow \cdots$ order we used in the Hamiltonian ansatz (\cref{sect:sim_scheme}). Note here we only talk about how to arrange the orbitals to have the right spin ordering, for the ordering of the orbitals within the same spin, it is determined by order of the rows and columns of the Slater determinants that we wrote down. In total, we need 
\begin{align*}
&\quad \frac{(\frac{N_{orb}}{2} - 2 + 1)\left(\frac{N_{orb}}{2} - 2\right)}{2} + \lfloor\frac{N_{orb}}{4}\rfloor \\
&\approx \frac{N_{orb}^2}{8} - \frac{N_{orb}}{2}\\
& = \frac{V^2}{2} - V
\end{align*}
fSWAPs to achieve the desired orbital order. 

For the fSWAP gates, similar to \cref{sect:gate_count_silicon}, we can arrange the gates such that the $\sqrt{Z}$ only acts on the odd orbitals, which will enable the cancellation of all the $\sqrt{Z}$ other than those at the boundary. Hence, we can ignore the  $\sqrt{Z}$ gates required by the fSWAPs. Our scheme need $\frac{V^2}{4} \times 2$ Givens rotations and $\frac{V^2}{2} - V$ fSWAPs. When decomposed into partial swap and $Z$ rotation, the number of one-qubit gates and two-qubit gates needed are
\begin{align*}
N_{1q} &= 2\left(\frac{V^2}{2}\right) + \frac{V^2}{2} - V = \frac{3}{2}V^2 - V\\
N_{2q} &= 2\left(\frac{V^2}{2}\right) + 2\left(\frac{V^2}{2} - V\right) = 2V^2 - 2V\\
\end{align*}

The depth of the circuit before finishing the first Givens rotation is $N_{orb} - N_e= V$. This section of the circuit consist of only Givens rotation, hence require runtime:
\begin{align*}
T_a = V\left(\tau_{1q} + 4\tau_{2q}\right)
\end{align*}
After this, we have the fSWAP network with Givens rotation happening concurrently, the depth of the circuit here is $V-1$, the runtime is limited by the fSWAP instead of Givens rotation since fSWAP contains $\pi$-rotations of $Z$. Hence, the runtime needed for the fSWAP networks is:
\begin{align*}
T_b = \left(V-1\right)\left(2\tau_{1q} + 4\tau_{2q}\right)
\end{align*}
Hence, the total runtime needed for the state preparation circuit is:
\begin{align*}
T = T_a + T_b = \left(3V -2\right)\tau_{1q} + \left(8V - 4\right)\tau_{2q}
\end{align*}
For $V = 25$, we have:
\begin{align*}
N_{1q} &= 910\\
N_{2q} &= 1250\\
T &= 73 \tau_{1q} + 196 \tau_{2q}
\end{align*}
Hence, the spin subspace scheme leads to some reduction in the number of one-qubit gate. However, it also leads to longer circuit runtime.

\subsubsection{Comparison to Ansatz circuit}
When compared to \cref{sect:gate_count_silicon}, the number of one-qubit gates of the Slater determinant preparation circuit is twice of that of one layer of the Hamiltonian ansatz, the number of two-qubit gates and the runtime are comparable.

Do note that the number of gates of the Slater determinant preparation circuit scale as $\mathcal{O}(V^2)$, which is worse than the $V^{3/2}$ ansatz scaling. The depth of the Slater determinant preparation circuit scale as $\mathcal{O}(V)$, which is also worse than the ansatz scaling $\mathcal{O}(\sqrt{V})$ as well. However, we need to note that we did not take into account of the number of blocks of Hamiltonian ansatz might be needed in the ansatz scaling, and the number of blocks needed is very likely to scale worse than $\sqrt{V}$, which means that the state preparation should have a better scaling than the ansatz if we take that into account.

\section{Gate Count Analysis for Superconducting Qubits}\label{sect:gate_count_super}
Now we will follow the same analysis in \cref{sect:ansatz_circuit} and \cref{sect:state_prep}, but switch to superconducting qubits. In superconducting qubits, we will use partial iSWAP~\cite{mckayUniversalGateFixedFrequency2016} instead of partial SWAP as our elementary two-qubit gate. The differences between the two are discussed in Ref.~\cite{schuchNaturalTwoqubitGate2003}

Note that partial iSWAP is \emph{not} $e^{-i \frac{\theta}{2}\text{i SWAP}}$, instead it is just a gate based on XY interaction:
\begin{align*}
\text{iSWAP}_{\theta} := e^{-i \frac{\theta}{2} \left(XX + YY\right)}\\
\text{iSWAP} := \text{iSWAP}_{-\pi/2}.
\end{align*}
In our circuit, partial iSWAP will be the only type of two-qubit gate we use. This would enable us to implement \emph{all} $Z$ rotations in a virtual way since the iSWAP can be easily transformed to implement virtual $Z$ gate~\cite{mckayEfficientZGatesQuantum2017}. Thus in the following gate count section we will omit all $Z$ rotations in our gate counts and time counts.

In the following section, we will use the below notations for the time units of the gates:
\begin{itemize}
    \item $\tau_{1q}$: the time unit for a single-qubit gate, which is the time needed to perform a $\frac{\pi}{2}$ rotation. We will assume the time needed to carry out a single-qubit gate with a variable parameter, i.e. gates like $X_\theta$, is on average $\tau_{1q}$.
    \item $\tau_{2q}$: the time unit for a two-qubit gate, which is the time needed to perform a iSWAP. We will assume the time needed to carry out a partial swap with a variable parameter, i.e. gates like $\text{iSWAP}_\theta$, is on average $\tau_{2q}$.
\end{itemize}
The decompositions of the gates in the ansatz into partial iSWAPs and single-qubit rotations and their resource estimates are shown below:
\begin{itemize}
    \item On-site repulsion: $U_{U}(\theta) = e^{-i \frac{\theta}{2} \left(I-Z_1\right)\left(I-Z_2\right)}$
    \begin{center}
        \begin{tikzcd}
            &\gate{Z_{\pi- \theta} }     &\gate[wires=2, style={yshift=-9.5pt, inner ysep=-5pt}]{\rotatebox{270}{i SWAP}}    &\qw &\gate[wires=2, style={yshift=-9.5pt, inner ysep=-5pt}]{\rotatebox{270}{i SWAP}}      &\gate{X_{\theta+ \pi}}&\gate[wires=2, style={yshift=-9.5pt, inner ysep=-5pt}]{\rotatebox{270}{i SWAP}}    &\qw &\gate[wires=2, style={yshift=-9.5pt, inner ysep=-5pt}]{\rotatebox{270}{i SWAP}} &\qw\\
            &\gate{Z_{\pi- \theta}} &     &\gate{X_{\frac{\pi}{2}}}&      &\qw 			   &     &\gate{X_{\frac{\pi}{2}}}&  &\qw   
        \end{tikzcd}
    \end{center}
    Gate counts: $G_{1q, U} = 3$, $G_{2q, U} = 4$.
    
    Time needed: $\tau_U = 5\tau_{1q} + 4\tau_{2q}$
    
    \item Hopping interaction: $U_{t}(\theta) = e^{-i \frac{\theta}{2} (XX + YY)} = \text{iSWAP}_{\theta}$
    
    Gate counts: $G_{1q, t} = 0$, $G_{2q, t} = 1$
    
    Time needed: $\tau_{t} = \tau_{2q}$
    \item Fermionic swap: \\$F_{sw} = \frac{1}{2} \left(XX + YY + ZI + IZ\right)$
    \begin{center}
        \begin{tikzcd}
            &\gate[wires=2, style={yshift=-9.5pt, inner ysep=-5pt}]{\rotatebox{270}{i SWAP}}    &\gate{Z_{-\frac{\pi}{2}}} &\qw\\
            &  &\gate{Z_{-\frac{\pi}{2}}}&\qw
        \end{tikzcd}
    \end{center}
    Gate counts: $G_{1q, F} = 0$, $G_{2q, F} = 1$
    
    Time needed: $\tau_f = \tau_{2q}$
    \item Fermionic swap + on-site repulsion: $F_{sw}U_{U}$
    
    Using the fact that iSWAP commute with $Z_{\theta}\otimes Z_{\theta}$ and $\text{iSWAP} \cdot \text{iSWAP} = Z\otimes Z$, we have:
    \begin{center}
        \begin{tikzcd}
            &\gate{Z_{-\frac{\pi}{2}- \theta} }     &\qw &\gate[wires=2, style={yshift=-9.5pt, inner ysep=-5pt}]{\rotatebox{270}{i SWAP}}      &\gate{X_{\theta+ \pi}}&\gate[wires=2, style={yshift=-9.5pt, inner ysep=-5pt}]{\rotatebox{270}{i SWAP}}    &\qw &\gate[wires=2, style={yshift=-9.5pt, inner ysep=-5pt}]{\rotatebox{270}{i SWAP}} &\qw\\
            &\gate{Z_{-\frac{\pi}{2}- \theta}} &\gate{X_{\frac{\pi}{2}}}&      &\qw 			   &     &\gate{X_{\frac{\pi}{2}}}&  &\qw   
        \end{tikzcd}
    \end{center}
    
    Gate counts: $G_{1q, FU} = 3$, $G_{2q, FU} = 3$.
    
    Time needed: $\tau_{FU}  = 5\tau_{1q} + 3\tau_{2q}$
    \item Fermionic swap + hopping interaction: $F_{sw}U_{t}$
    \begin{center}
        \begin{tikzcd}
            &\gate[wires=2, style={yshift=-13.5pt, inner ysep=-5pt}]{\rotatebox{270}{$\text{i SWAP}_{\theta - \frac{\pi}{2}}$}}    &\gate{Z_{-\frac{\pi}{2}}} &\qw\\
            &  &\gate{Z_{-\frac{\pi}{2}}}&\qw
        \end{tikzcd}
    \end{center}
    Gate counts: $G_{1q, Ft} = 0$, $G_{2q, Ft} = 1$
    
    Time needed: $\tau_{Ft} = 2\tau_{2q}$
\end{itemize}
Following the above gate decomposition and the Hamiltonian ansatz circuit outlined in \cref{sect:ansatz_circuit}, we can obtain the following estimates for the total number of one-qubit gates needed $N_{1q, antz}$ (all are $X$ rotations), total number of two-qubit gates needed $N_{2q, antz}$ (all are partial swaps) and the total length of time needed $T_{antz}$ to perform one block of Hamiltonian ansatz for 2D Hubbard model of $V$ sites using basic single-qubit rotation and partial iSWAPs:
\begin{align*}
N_{1q, antz} & = 3V\\
N_{2q, antz} & = 4V^{\frac{3}{2}} + 2V - 2\sqrt{V}\\
T_{antz} &= 5\tau_{1q} + 4\left(\sqrt{V} + 1\right) \tau_{2q} 
\end{align*}

For $V = 25$, we have:
\begin{align*}
N_{1q, antz} & \approx 75\\
N_{2q,antz} & \approx 540\\
T_{antz} & \approx 5 \tau^{(1)} + 24 \tau^{(2)}
\end{align*}

Now we will turn to initial Slater determinant preparation outlined in \cref{sect:state_prep}. The gate needed is Given rotation, which can be decomposed as
\begin{itemize}
    \item Givens rotation: $R(\theta) = e^{-i\frac{\theta}{2}(YX - XY)}$
    \begin{center}
        \begin{tikzcd}
            &\qw&\gate[wires=2, style={yshift=-10.5pt, inner ysep=-5pt}]{\rotatebox{270}{$\text{i SWAP}_\theta$}}    &\qw &\qw\\
            &\gate{Z_{\frac{\pi}{2}}}&  &\gate{Z_{-\frac{\pi}{2}}}&\qw
        \end{tikzcd}
    \end{center}
    Gate counts: $G_{1q, G} = 0$, $G_{2q, G} = 1$
    
    Time needed: $\tau_{G} = \tau_{2q}$.
\end{itemize}
Following the same arguments in \cref{sect:gate_count_state_prep}, the number of two-qubit gates and one-qubit gates needed for Slater determinant preparation is: 
\begin{align*}
N_{1q,prep} &= V^2\\
N_{2q,prep} &= V^2.
\end{align*}
The depth of the Slater determinant preparation circuit is
\begin{align*}
D =  2V-1
\end{align*}
which translate into the circuit runtime of
\begin{align*}
T_{prep} = \left(2V-1\right) \tau_{2q}.
\end{align*}

Together we can obtain the resource estimate for the full ansatz circuit with $N_{blk}$ ansatz blocks:
\begin{align*}
N_{1q} & = V^2 + 3V N_{blk} \\
N_{2q} & = V^2 + \left(4V^{\frac{3}{2}} + 2V - 2\sqrt{V}\right)N_{blk}\\
T &= \left(2V-1\right) \tau_{2q} + \left(5\tau_{1q} + 4\left(\sqrt{V} + 1\right) \tau_{2q} \right) N_{blk} 
\end{align*}
For $V = N_{blk} = 25$, we have:
\begin{align*}
N_{1q} & \approx 2500 \\
N_{2q} & \approx 14000\\
T &\approx  125 \tau_{1q} +  650\tau_{2q}.
\end{align*}

\section{Obtaining Energy Gradient in Quantum Computers}\label{sect:grad_method}
\subsection{Background}
As mentioned in \cref{sect:vqe}. For a given Hamiltonian $H$, we want to find the set of optimal parameters $\vec{\theta}$ for an ansatz circuit such that the state $\ket{\psi(\vec{\theta})}$ it produce is as close to the ground state of the Hamiltonian as possible, i.e. we want to find $\vec{\theta}$ such that $E_{tot}(\vec{\theta}) = \bra{\psi(\vec{\theta})} H \ket{\psi(\vec{\theta})}$ is minimised.

However, since the energy $E_{tot}(\vec{\theta})$ cannot be directly measured, we need to rewrite our Hamiltonian in terms of its Pauli components:
\begin{align*}
H &= \sum_j \lambda_j G_j\\
\bra{\psi(\vec{\theta})} H \ket{\psi(\vec{\theta})} &=  \sum_j \lambda_j \bra{\psi(\vec{\theta})} G_j \ket{\psi(\vec{\theta})}.
\end{align*}
Hence,
\begin{align*}
E_{tot}(\vec{\theta})& =  \sum_j \lambda_j E_{i}(\vec{\theta}) 
\end{align*}
where $E_j(\vec{\theta}) = \bra{\psi(\vec{\theta})} G_j \ket{\psi(\vec{\theta})}$, which is just the expectation value of a Pauli observable, hence can be directly obtained from the circuit.

To find the ground state, we can use gradient-based optimisation methods like gradient decent and L-BFGS-B, etc. This require us to obtain $\pdv{E_{tot}(\vec{\theta})}{\theta_m}$, which in turns means that we need to measure $\pdv{E_{j}(\vec{\theta})}{\theta_m}$ for every Hamiltonian component $j$ and every parameter $m$.

We can obtain $\pdv{E_{j}(\vec{\theta})}{\theta_m}$ simply using finite difference, in which we will measure $E_{j}(\theta_1, \theta_2, \cdots \theta_m, \cdots)$ and $E_{j}(\theta_1, \theta_2, \cdots \theta_m + \delta \theta_m, \cdots)$, and take their difference divided by $\delta \theta_m$ to approximate the gradient.

\subsection{The circuit approach to obtain gradients}\label{sect:circuit_grad_meas}
The exact gradient $\pdv{E_{j}(\vec{\theta})}{\theta_m}$ can be obtained using circuit measurement instead of using finite difference approximation. Suppose our ansatz circuit consist of parametrised rotation gates $R_n$ of the form
\begin{align*}
R_n(\theta_n) &= e^{-i\theta_n F_n}
\end{align*}
where $F_n$ are both Hermitian and unitary, e.g. Pauli or swaps. 

Then our parametrised circuit can be written as:
\begin{align*}
\ket{\psi(\vec{\theta})} = \prod_{n = N}^{1} R_n(\theta_n) \ket{0} = R(\vec{\theta}) \ket{0}
\end{align*}
where we have denote the whole circuit using $R(\vec{\theta})$.

A string of parametrised rotation can be denoted as:
\begin{align*}
\prod_{n = a}^{b} R_n(\theta_n) = R_{a:b}
\end{align*}
which means that $R(\vec{\theta}) = R_{N:1}$

\begin{figure}[htbp]
    \centering
    \subfloat[]{\includegraphics[width = 0.5\textwidth]{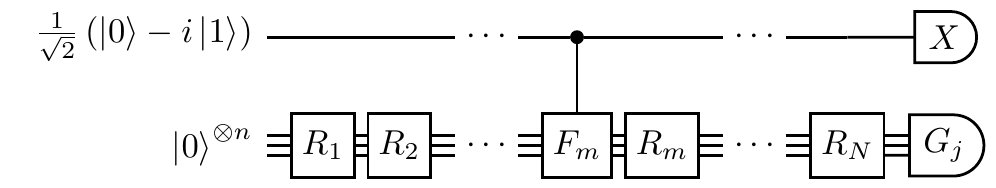}}\\
    \subfloat[]{\includegraphics[width = 0.5\textwidth]{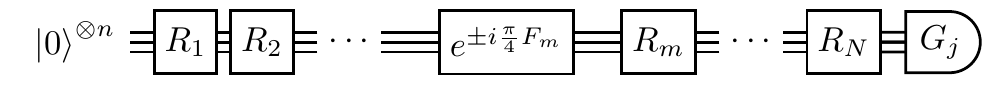}}
    \caption{These are the circuits used to measure the gradient $\pdv{E_{j}(\vec{\theta})}{\theta_m}$ using (a) indirect measurements, (b) direct measurements. Here $E_{j}(\vec{\theta}) = \bra{\psi(\vec{\theta})} G_j \ket{\psi(\vec{\theta})}$ is the expectation value of the $j$th component of the Hamiltonian. $\theta_m$ is the parameter of the $m$th parametrised gate $R_m(\theta_m) = e^{-i\theta_m F_m}$ of the ansatz circuit, where $F_m$ is both Hermitian and unitary. Note that here we have assumed that all parametrised gates have independent parameters.}
    \label{fig:grad_circ}
\end{figure}

Hence, 
\begin{align*}
\pdv{R(\vec{\theta})}{\theta_m} &=  R_{N:m+1} \pdv{R_m(\theta_m)}{\theta_m} R_{m-1:1}\\
& = -i R_{N:m+1} F_m R_{m:1}\\
\end{align*}
Hence we have
\begin{align*}
\pdv{E_{j}(\vec{\theta})}{\theta_m} &= 2\Re{\bra{\psi(\vec{\theta})}G_j\left(\partial_m\ket{\psi(\vec{\theta})}\right)}\\
& = 2\Re{\bra{\overline{0}} R^\dagger(\vec{\theta})G_j\left(\partial_mR(\vec{\theta})\right)\ket{\overline{0}}}\\
& = 2\Im{\bra{\overline{0}} \left(R_{N:1}\right)^\dagger G_jR_{N:m+1} F_m R_{m:1}\ket{\overline{0}}}
\end{align*}
This can be measured via two kind of circuits:
\begin{itemize}
    \item Indirect measurement~\cite{liEfficientVariationalQuantum2017}:
    
    The circuit is shown in \cref{fig:grad_circ}~(a) will measure $\frac{1}{2}\pdv{E_{j}(\vec{\theta})}{\theta_m}$. Here we use an ancilla which probes the main ansatz with a control unitary. The advantage of such a scheme is that we can obtain the exact gradient of the circuit via measurement of only one qubit. However, it requires an extra ancilla qubit. The need for performing the control unitary between the ancilla and any other qubits also lead to connectivity challenges. Control unitaries like control swap are also not straightforward to implement in lots of architectures.
    
    \item Direct measurement~\cite{mitaraiMethodologyReplacingIndirect2019}:
    
    The circuit is shown in \cref{fig:grad_circ}~(b). 
    Denoting $\ket{\phi} = R_{m:1}\ket{\overline{0}}$, $W = R_{N:m+1}$, the circuit will measure:
    \begin{align*}
    A_{jm, \pm}& = \bra{\phi} e^{ \pm i\frac{\pi}{4}F_m} W^\dagger G_jW e^{\mp i\frac{\pi}{4}F_m} \ket{\phi}\\
    & = \frac{1}{2}\bra{\phi}\left(1 \pm iF_m\right) W^\dagger G_jW \left(1 \mp iF_m\right) \ket{\phi}\\
    & = \frac{1}{2}\bra{\phi}W^\dagger G_jW \ket{\phi} + \frac{1}{2}\bra{\phi}F_m W^\dagger G_jW F_m \ket{\phi}\\
    & \ \pm i \frac{1}{2}\left(\bra{\phi}F_m W^\dagger G_jW \ket{\phi} - \bra{\phi} W^\dagger G_jW F_m \ket{\phi}\right)
    \end{align*}
    Hence, the gradient $\pdv{E_{j}(\vec{\theta})}{\theta_m}$ can be obtained via 
    \begin{align*}
    &\quad A_{jm, +} - A_{jm, -} \\
    & = i \left(\bra{\phi}F_m W^\dagger G_jW \ket{\phi} - \bra{\phi} W^\dagger G_jW F_m \ket{\phi}\right) \\
    &= 2\Im{\bra{\phi} W^\dagger G_jW F_m \ket{\phi}}\\
    & = \pdv{E_{j}(\vec{\theta})}{\theta_m}
    \end{align*}
    For this method, we do not require any extra ancilla or control unitaries. However, we need to measure two expectation values $A_{jm, +}$ and $A_{jm, -}$ for the estimation of the gradient $\pdv{E_{j}(\vec{\theta})}{\theta_m}$ instead of one in the case of indirect measurement.
\end{itemize}
\subsection{Shared parameters}\label{sect:shared_para}
For the case where there are parametrised gates with shared parameters, we can still obtain the gradients using finite difference in a similar way.

When using gradient circuits, the story is more complicated. Firstly, we add a scaling factor $\beta_m$ to each parameters: $\theta_m \rightarrow \beta_m \theta_m$:
\begin{align*}
\pdv{R(\{\theta_n\})}{\theta_m} &= -i R_{N:m+1} F_m R_{m:1} \\
\Rightarrow \pdv{R(\{\beta_n\theta_n\})}{\beta_m\theta_m} &= -i R_{N:m+1} F_m R_{m:1}\\
\pdv{R(\{\beta_n\theta_n\})}{\theta_m} &= -i \beta_m R_{N:m+1} F_m R_{m:1}\\
\end{align*}
Now we change the labelling of the parametrised gates: $m \rightarrow m, v$, for which gates with the same $m$ will share the same parameters, and $v$ labels the different parametrised gates that share the same parameter:
\begin{align*}
\theta_{m,v} = \theta_m\quad \forall v
\end{align*}
In such case, we have 
\begin{align*}
\pdv{R(\{\beta_{n, w}\theta_{n, w}\})}{\theta_m} &= \sum_{v} \left[\pdv{R(\{\beta_{n, w}\theta_{n, w}\})}{\theta_{m, v}}\right]_{\theta_{m, v} = \theta_m \forall v} 
\end{align*}
i.e. the gradient w.r.t. to a given parameter $\theta_m$ is the sum of all the gradient w.r.t. the parameter of each parametrised gate that shared the parameter value. Hence, when trying to obtain the gradient using circuit measurements, we still need to treat the parameters in each gate as independent, and then sum those gradient up based on which gates have shared parameters. 

\section{Number of Samples Needed for Energy and Energy Gradient}\label{sect:grad_n_sample}
\subsection{Number of samples in finite difference}
\subsubsection{Gradient precision}
The equation for the estimation of the gradient of the $j^{th}$ Pauli term in the Hamiltonian using finite difference is:
\begin{align}\label{eqn:finite_diff}
\pdv{E_{j}(\vec{\theta})}{\theta_m} &= \frac{\overline{E}_{j}(\vec{\theta} + \frac{\vec{\delta}_m}{2})  - \overline{E}_{j}(\vec{\theta} - \frac{\vec{\delta}_m}{2})}{\delta}
\end{align}
where $\vec{\delta}_m$ is a vector with the $m^{th}$ parameter set to $\delta$ and all other parameters set to $0$. $\overline{E}_{j}$ denote the sampling average of $E_{j}$. 

Hence, we have
\begin{align}\label{eqn:var_grad_fd_1}
\var{\partial_mE_j} = \frac{2}{\delta^2} \var{\overline{E}_j}
\end{align}
i.e. for larger $\delta$, we can achieve smaller variance in gradient for a fixed variance in energy. However, we cannot increase $\delta$ indefinitely because there is an error associated with the finite step size when using finite difference, which has the magnitude of:
\begin{align*}
\frac{\delta^2}{24} \sum_{u,v,w = 1}^{N_{sh}} \pdv{^3E_{j}(\vec{\theta})}{\theta_{m, v}\partial\theta_{m, u}\partial\theta_{m, w}} \approx N_{sh}^3\frac{\delta^2}{24}  \partial^3_m E_{j}
\end{align*}
where number of parametrised gates share parameter $\theta_{m}$ is $N_{sh}$. The sum over $v, u, w$ are the sum over all the parametrised gates that share the same parameter $\theta_m$ as discussed in \cref{sect:shared_para}. We have made the assumption that all third-order derivatives in the sum have similar magnitudes and the same sign.  Note that the finite step size error increases with $\delta$.

A balance between the two errors can be achieved when we choose the step size $\delta$ to satisfy
\begin{align*}
\sqrt{\frac{2}{\delta^2} \var{\overline{E}_j}} &= N_{sh}^3 \frac{\delta^2}{24} \partial^3_m E_{j}\\
\delta & = \left(\frac{24\sqrt{2\var{\overline{E}_j}}}{N_{sh}^3\partial_m^3 E_j}\right)^{\frac{1}{3}}
\end{align*}
Note that 
\begin{align*}
\pdv{^3E_{j}(\vec{\theta})}{\theta_m^3} &= 2\Re{\bra{\psi(\vec{\theta})}G_j\left(\partial_m^3\ket{\psi(\vec{\theta})}\right)} \\
& \quad + 6\Re{\left(\partial_m\bra{\psi(\vec{\theta})}\right)G_j\left(\partial_m^2\ket{\psi(\vec{\theta})}\right)}
\end{align*}
where $\Re{\left(\partial_m\bra{\psi(\vec{\theta})}\right)G_j\left(\partial_m^2\ket{\psi(\vec{\theta})}\right)}$ and $\Re{\bra{\psi(\vec{\theta})}G_j\left(\partial_m^3\ket{\psi(\vec{\theta})}\right)}$ can be measured using circuit similar to the first order derivative in \cref{sect:circuit_grad_meas}. Hence, the magnitude of $\pdv{^3E_{j}(\vec{\theta})}{\theta_m^3}$ is around $ \sqrt{2^2 + 6^2} \approx 6$.

Thus we have:
\begin{align*}
\delta & = \left(\frac{24\sqrt{2\var{\overline{E}_j}}}{6N_{sh}^3}\right)^{\frac{1}{3}}\\
& = \frac{1.78}{N_{sh}} \var{\overline{E}_j}^{\frac{1}{6}}
\end{align*}
Substituting into \cref{eqn:var_grad_fd_1}, we have:
\begin{equation}\label{eqn:var_grad_fd}
    \begin{split}
        \var{\partial_mE_j} &= 0.63N_{sh}^2 \var{\overline{E}_j}^{\frac{2}{3}}\\
        \var{\overline{E}_j} &= \frac{2}{N_{sh}^3}\var{\partial_mE_j}^{\frac{3}{2}}
    \end{split}
\end{equation}
This is the smallest variance in the gradient that we can achieve for a given variance in the energy estimation by choosing the optimal step size $\delta$.
\subsubsection{Number of samples needed}
Assuming $\var{E_{j}} \sim \mathcal{O}(1)$, then the number of samples needed to achieve the sample average variance $\var{\overline{E}_j}$ is
\begin{align*}
\frac{\var{E_{j}}}{\var{\overline{E}_j}} \sim \frac{1}{\var{\overline{E}_j}}
\end{align*}
Since we need to evaluate sample average $\overline{E}_j$ at two points in finite difference, the total number of samples needed is:
\begin{align}\label{eqn:samples_fd_one}
M_{fd} \sim \frac{2}{\var{\overline{E}_j}} = \frac{N_{sh}^3}{\var{\partial_mE_j}^{\frac{3}{2}}}
\end{align}

\subsection{Number of samples in direct measurement}
\subsubsection{Gradient precision}
The equation of estimation of the gradient of a the $j^{th}$ Pauli term in the Hamiltonian using direct measurement is:
\begin{align*}
\pdv{E_{j}(\vec{\theta})}{\theta_m} &= \sum_{v = 1}^{N_{sh}} \left[\pdv{E_{j}(\vec{\theta})}{\theta_{m, v}}\right]_{\theta_{m,v} = \theta_m \forall v} \\
& = \sum_{v=1}^{N_{sh}} \left(\overline{A}_{jmv, +} - \overline{A}_{jmv, -} \right)
\end{align*}
Here the sum over $v$ is the sum over all the parametrised gates that share the same parameter $\theta_m$ as discussed in \cref{sect:shared_para}.

Hence, we have:
\begin{align*}
\var{\partial_mE_j} = 2N_{sh} \var{\overline{A}}
\end{align*}

\subsubsection{Number of samples needed}
Assuming $\var{A} \sim \mathcal{O}(1)$, then the number of samples needed to achieve sample average variance $\var{\overline{A}}$ is
\begin{align*}
\frac{\var{A}}{\var{\overline{A}}} \sim \frac{1}{\var{\overline{A}}}
\end{align*}
Since we need to evaluate sample average $\overline{A}$ at $2N_{sh}$ points in direct measurement, the total number of samples needed is:
\begin{align}\label{eqn:samples_dm_one}
M_{dm} \sim \frac{2N_{sh}}{\var{\overline{A}}} = \frac{\left(2N_{sh}\right)^2}{\var{\partial_mE_j}}
\end{align}
When compared to \cref{eqn:samples_fd_one} of finite difference, we can see that direct measurement has better scaling in terms of both the number of shared of parameters $N_{sh}$ and the target gradient precision $\var{\partial_mE_j}$, thus direct measurement is preferred.

\subsection{Comparison between finite difference and direct measurement}\label{sect:fd_dm_comp}
To compare the two methods, we use \cref{eqn:samples_fd_one} and \cref{eqn:samples_dm_one} to study
\begin{align*}
\frac{M_{dm}}{M_{fd}} = \frac{4 \sqrt{\var{\partial_mE_j}}}{N_{sh}}
\end{align*}
If we define the breaking point of the gradient precision as:
\begin{align}
\epsilon_{grad}^* = \frac{N_{sh}}{4}
\end{align}
then we have:
\begin{itemize}
    \item $\sqrt{\var{\partial_mE_j}} \geq \epsilon_{grad}^* \Rightarrow \frac{M_{dm}}{M_{fd}} \geq 1$:
    
    Finite difference need less samples to achieve the given precision.
    \item $\sqrt{\var{\partial_mE_j}} < \epsilon_{grad}^* \Rightarrow \frac{M_{dm}}{M_{fd}} < 1$:
    
    Direct measurement need less samples to achieve the given precision.
\end{itemize}

$N_{sh} = 4N_{eq}$ for hopping term (if we assume spin symmetry) and $N_{sh} = 3N_{eq}$ for repulsion term where $N_{eq}$ is the number of equivalent partition in the site layout due to symmetry. Here we will take the approximation that all $N_{sh} = 4N_{eq}$ since there are more hopping term than repulsion terms. Thus we have:
\begin{align}\label{eqn:var_grad_dm}
\var{\partial_mE_j} = 8N_{eq} \var{\overline{A}}
\end{align}

Using the parametrisation discuss in \cref{sect:parametrisation}, for open boundary Hubbard model the breaking point of the gradient precision $\epsilon_{grad}^*$ is:
\begin{itemize}
    \item Square site layout:
    \begin{align*}
    N_{eq} = 8 \Rightarrow \epsilon_{grad, sq}^* = 2.5 \times 10^{-4}
    \end{align*}
    \item Rectangular site layout:
    \begin{align*}
    N_{eq} = 4 \Rightarrow \epsilon_{grad, rt}^* = 1 \times 10^{-3}
    \end{align*}
\end{itemize}

\subsection{Number of samples needed for energy}\label{sect:energy_prec}
As discussed in \cite{corbozCompetingStatesModel2014}, to compete with the best classical algorithm we need to estimate the energy per site $E_{site}$ to $10^{-3} t$ precision. 

We can decompose the total energy into its subterms $E_j$:
\begin{align*}
E_{tot} \approx \sum_{j = 1}^{J} h_j E_j
\end{align*}
here $h_j$ is the coefficient of the Pauli decomposition of the Hamiltonian $H = \sum_{j = 1}^{J} h_j G_j$.

The repulsion terms and hopping terms in the Hamiltonian can be decomposed into their Pauli components:
\begin{align*}
E_{rep} &= \frac{1}{4} + \sum_{j = 1}^{3} \frac{1}{4} E_j\\
E_{hop} &= \sum_{j = 1}^{2} \frac{1}{2} E_j
\end{align*}
Hence, the variance in their sampling average are:
\begin{align*}
\var{\overline{E}_{rep}} &= \frac{3}{4^2}\var{\overline{E}_j}\\
\var{\overline{E}_{hop}} &= \frac{1}{2}\var{\overline{E}_j}
\end{align*}
There are $V$ repulsion terms and $4V$ hopping terms, hence the total energy is:
\begin{align*}
\sum_{k = 1}^{V} E_{rep, k} + \sum_{k = 1}^{4V} E_{hop, k} = VE_{site}
\end{align*}
which translate into the following equation for variance:
\begin{align*}
V \var{\overline{E}_{rep}} + 4V \var{\overline{E}_{hop}} &= V^2 \var{\overline{E}_{site}}\\
\var{\overline{E}_{j}} &= \frac{16}{35} V\var{\overline{E}_{site}}
\end{align*}
As mentioned above, we want to achieve $\var{\overline{E}_{site}} = \left(10^{-3}t\right)^2$, assuming $t \sim \mathcal{O}(1)$, we have:
\begin{align}\label{eqn:energy_prec}
\var{\overline{E}_{j}} &= \frac{16}{35} V \times 10^{-6}
\end{align}

Assuming $\var{E_{j}} \sim \mathcal{O}(1)$, then the number of samples needed to achieve sample average variance $\var{\overline{E}_j}$ is
\begin{align*}
M_{Ej} \sim \frac{\var{E_{j}}}{\var{\overline{E}_j}} \sim \frac{1}{\var{\overline{E}_j}}= \frac{35}{16V} \times 10^{6} .
\end{align*}
As mentioned in \cref{sect:energy_measurement}, we need 5 circuit runs to evaluate all energy subterms. Thus, the total number of circuit runs needed to evaluate the energy to the required precision is:
\begin{align*}
M_{E} = 5 M_{Ej} = \frac{1.1\times 10^{7}}{V} 
\end{align*}
Thus for $V = 25$, we have:
\begin{align*}
M_{E} \approx 4 \times 10^{5}
\end{align*}
\subsection{Number of samples needed for energy gradients}\label{sect:grad_prec}
First we need to decide what precision of the energy gradient is needed. In \cref{sect:energy_prec}, we have obtained the precision of the energy subterms that we want to achieve. If we are using finite difference method, in order to achieve such a precision in the final result energy subterms, we can evaluate the energy points in the gradient estimation to the same precision, and our terminating threshold of change in the estimated energy subterms can be set to the same precision. In such case, the gradient precision that we can achieve can be obtained using \cref{eqn:var_grad_fd} and \cref{eqn:energy_prec}:
\begin{align}\label{eqn:grad_prec}
\var{\partial_mE_j} & = 0.63 N_{sh}^2\var{\overline{E}_j}^{\frac{2}{3}} \nonumber\\
& = 3.7N_{sh}^2V^{\frac{2}{3}} \times 10^{-5}
\end{align}
To achieve the same precision using direct measurement, the number of sample needed can be obtained by substituting this into \cref{eqn:samples_dm_one}:
\begin{align}
M_{dm} \approx \frac{\left(2N_{sh}\right)^2}{\var{\partial_mE_j}} = 1.1 V^{-\frac{2}{3}} \times 10^{5}.
\end{align}

Note that this is just the number of samples needed to obtain the gradient of an energy subterm w.r.t. \emph{one} parameters. To obtain the full gradient vector, we need to iterate over all parameters. Using \cref{eqn:n_para}, the number of circuit runs needed to evaluate $\pdv{E_{j}(\vec{\theta})}{\theta_m}$ for all parameters $m$ using direct measurement is 
\begin{align*}
M_{grad,j} &= M_{dm}N_{para} \\
&\approx \frac{1.1 N_{para}^{site} N_{blk}V^{\frac{1}{3}}}{N_{eq}} \times 10^{5}
\end{align*}

As mentioned in \cref{sect:energy_measurement}, we need 5 circuit runs to measure all energy subterms. Thus, the total number of circuit runs needed to evaluate the energy gradient vector to the required precision is:
\begin{align*}
M_{grad} = 5 M_{grad,j} = \frac{5.5 N_{para}^{site} N_{blk}V^{\frac{1}{3}}}{N_{eq}} \times 10^{5}
\end{align*}

 For $5 \times 5$ Hubbard model, we have $V=25$, $N_{eq} = 8$ and $N_{para}^{site} = 5$. Assuming $N_{blk} = V$, the number of circuit runs needed is:
\begin{align*}
M_{grad} \approx 2.5 \times 10^{7}
\end{align*}

\section{Quantum Dot Layout}\label{sect:dot_layout}
\begin{figure}[h!]
    \centering
    \includegraphics[width = 0.5\textwidth]{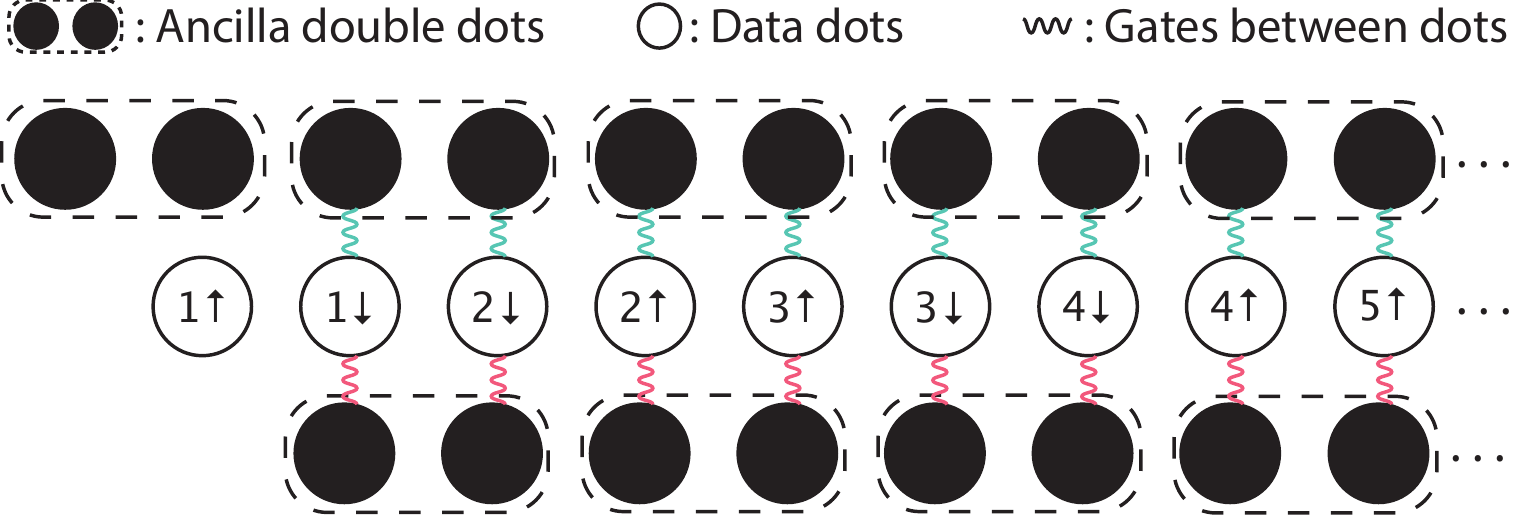}
    \caption{The measurements of the hopping terms. The blue interaction here are control-$X$ gate from the ancilla to the data for the measurement of $XX$ in the hopping term. The red interaction here are control-$Y$ gate from the ancilla to the data for the measurement of $YY$ in the hopping term.}
   \label{fig:arch_hop}
\end{figure}

\begin{figure}[h!]
    \centering
    \includegraphics[width = 0.5\textwidth]{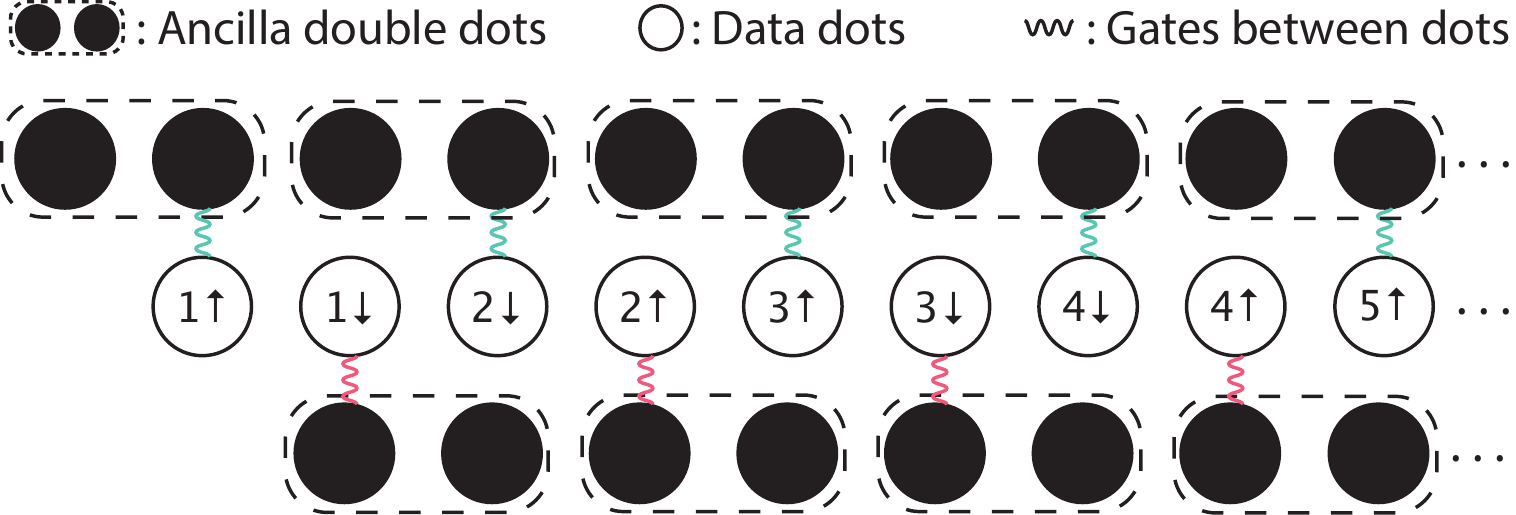}
    \caption{The measurements of the repulsion terms. All the interaction here are control-$Z$ gates. One row of ancilla is in charge of $Z$ measurements of the odd data dots while the other row of ancilla is in charge of $Z$ measurements of the even data dots.}
    \label{fig:arch_rep}
\end{figure}

\cref{fig:arch_hop}, \cref{fig:arch_rep} shows how we can perform the non-demolishing measurements required by our problem using three lines of quantum dot, one line for data and two lines for ancilla. The ancilla here come in the form of double dot, which will be initialised in singlet and will be readout using Pauli spin blockade which enable us to distinguish singlet and triplet state. \cref{fig:arch_hop} shows how we can measure $XX$ and $YY$ of the hopping terms. After that, we can compose $XX$ and $YY$ to obtain $ZZ$ (with an additional $-$ sign) which in terms can give us electron number parity for symmetry verification. Alternatively we can also measure $XX$ and $ZZ$, and obtain $YY$ via post-processing since control-$Z$ gate can be easier to implement in silicon qubits. \cref{fig:arch_hop} shows how we perform $Z$-measurement for every data dots to obtain the repulsion terms. The electron number parity in symmetry verification can again be obtained via composing our measurement results. Our measurement can also be carried out using only one row of ancilla. In such a case, we essentially using one row of ancilla to carry out two rounds of measurements with ancilla reinitialisation in between. Other architectures that incorporating spin-to-charge readout using reservoirs are also possible.
\FloatBarrier

%

\end{document}